\documentclass[acmtog,authorversion,nonacm]{acmart}

\usepackage{booktabs} 

\usepackage[ruled]{algorithm2e} 

\SetAlFnt{\small}
\SetAlCapFnt{\small}
\SetAlCapNameFnt{\small}
\SetAlCapHSkip{0pt}

\usepackage{subfigure}
\usepackage{caption}
\usepackage{enumitem}
\SetKwComment{Comment}{$\triangleright$\ }{}

\usepackage{color}
\usepackage{xcolor}

\usepackage{multirow}

\begin{document}

\title{F-RDW: Redirected Walking with Forecasting Future Position}

\author{Sang-Bin Jeon}
\email{ludens0508@yonsei.ac.kr}
\orcid{0000-0003-0388-3171}
\author{Jaeho Jung}
\email{gkgkgkwogh@yonsei.ac.kr}
\orcid{}
\author{Jinhyung Park}
\email{jh9604@yonsei.ac.kr}
\orcid{0000-0002-6273-6182}
\author{In-Kwon Lee}
\email{iklee@yonsei.ac.kr}
\orcid{0000-0002-1534-1882}
\affiliation
{
 \institution{Dept. of Computer Science, Yonsei University}
 \streetaddress{Yonsei-Ro 50}
 \city{Seodaemungu}
 \state{Seoul}
 \country{Rep. of Korea}
 \postcode{03722}
}

\renewcommand{\shortauthors}{Jeon et al.}

\begin{abstract}
    In order to serve better VR experiences to users, existing predictive methods of Redirected Walking (RDW) exploit future information to reduce the number of reset occurrences. However, such methods often impose a precondition during deployment, either in the virtual environment's layout or the user's walking direction, which constrains its universal applications. 
    To tackle this challenge, we propose a novel mechanism \textit{F-RDW} that is twofold: (1) forecasts the future information of a user in the virtual space without any assumptions, and (2) fuse this information while maneuvering existing RDW methods.  
    The backbone of the first step is an LSTM-based model that ingests the user's spatial and eye-tracking data to predict the user's future position in the virtual space, and the following step feeds those predicted values into existing RDW methods (such as MPCRed, S2C, TAPF, and ARC) while respecting their internal mechanism in applicable ways.
    The results of our simulation test and user study demonstrate the significance of future information when using RDW in small physical spaces or complex environments. 
    We prove that the proposed mechanism significantly reduces the number of resets and increases the traveled distance between resets, hence augmenting the redirection performance of all RDW methods explored in this work.
\end{abstract}


    \begin{teaserfigure}
        \centering
            \subfigure[\label{fig:ARC_VE}]
            {\includegraphics[width=0.35\textwidth]{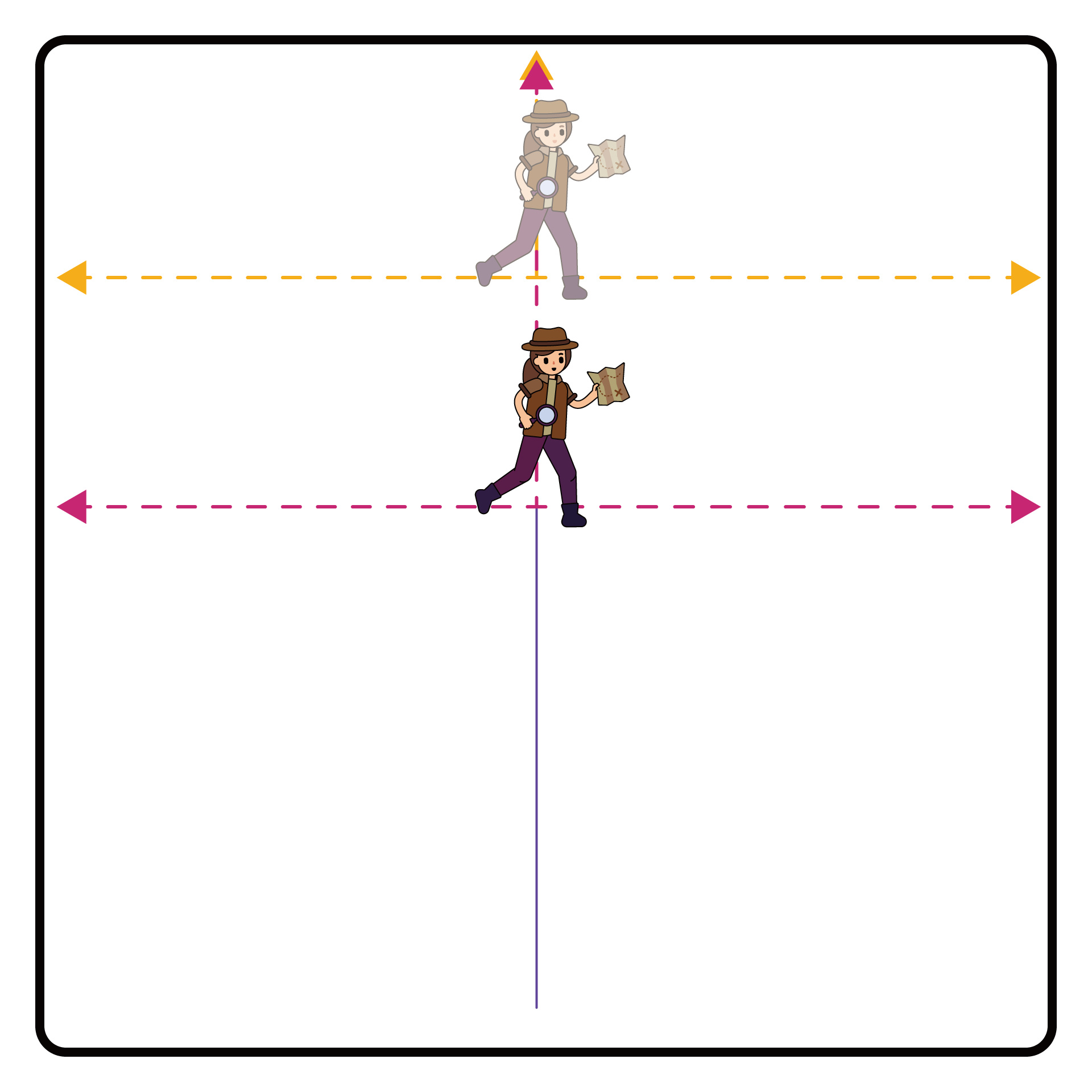}}
            \hfill
            \subfigure[\label{fig:ARC_PE}]
            {\includegraphics[width=0.3\textwidth]{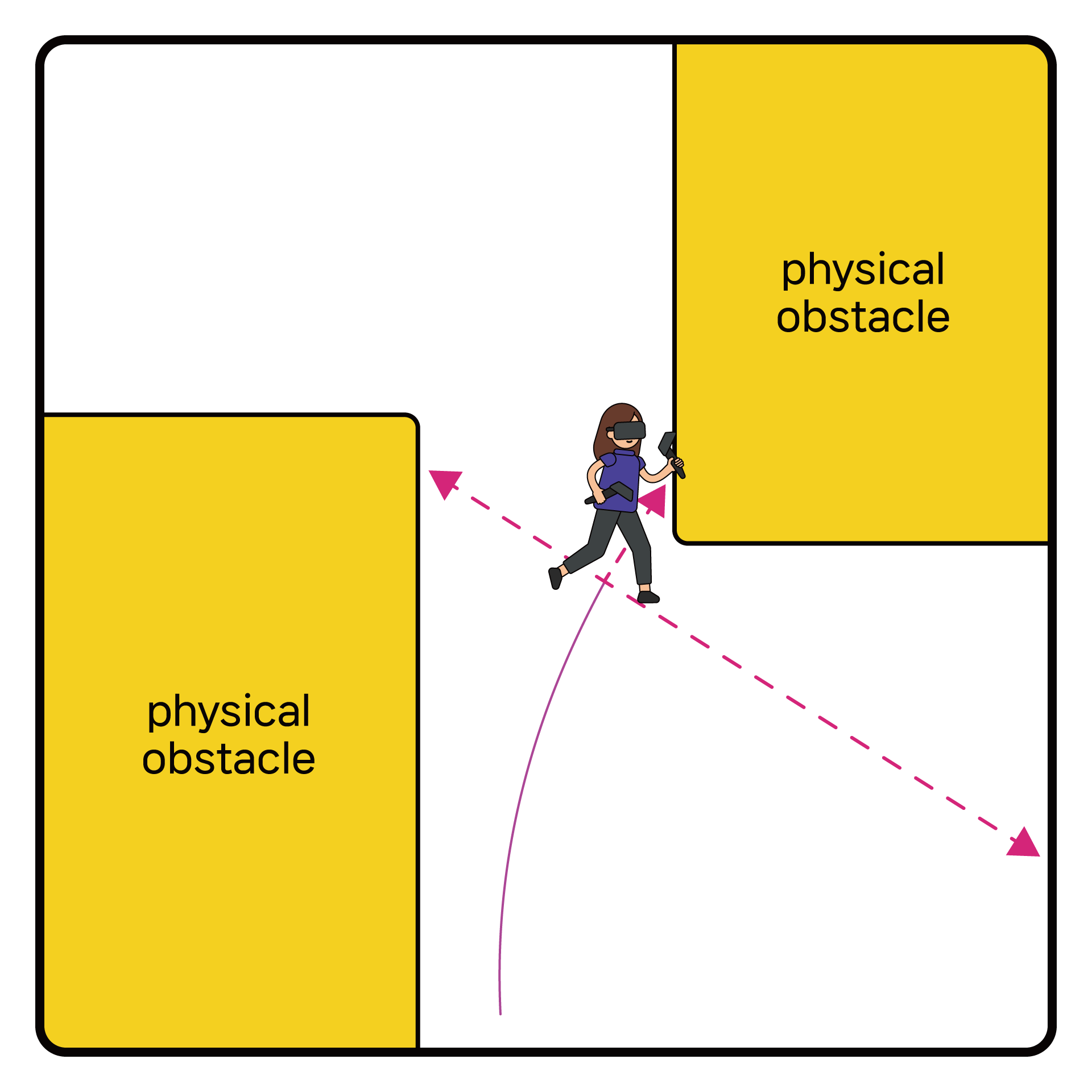}}
            \hfill
            \subfigure[\label{fig:F-ARC_PE}]
            {\includegraphics[width=0.3\textwidth]{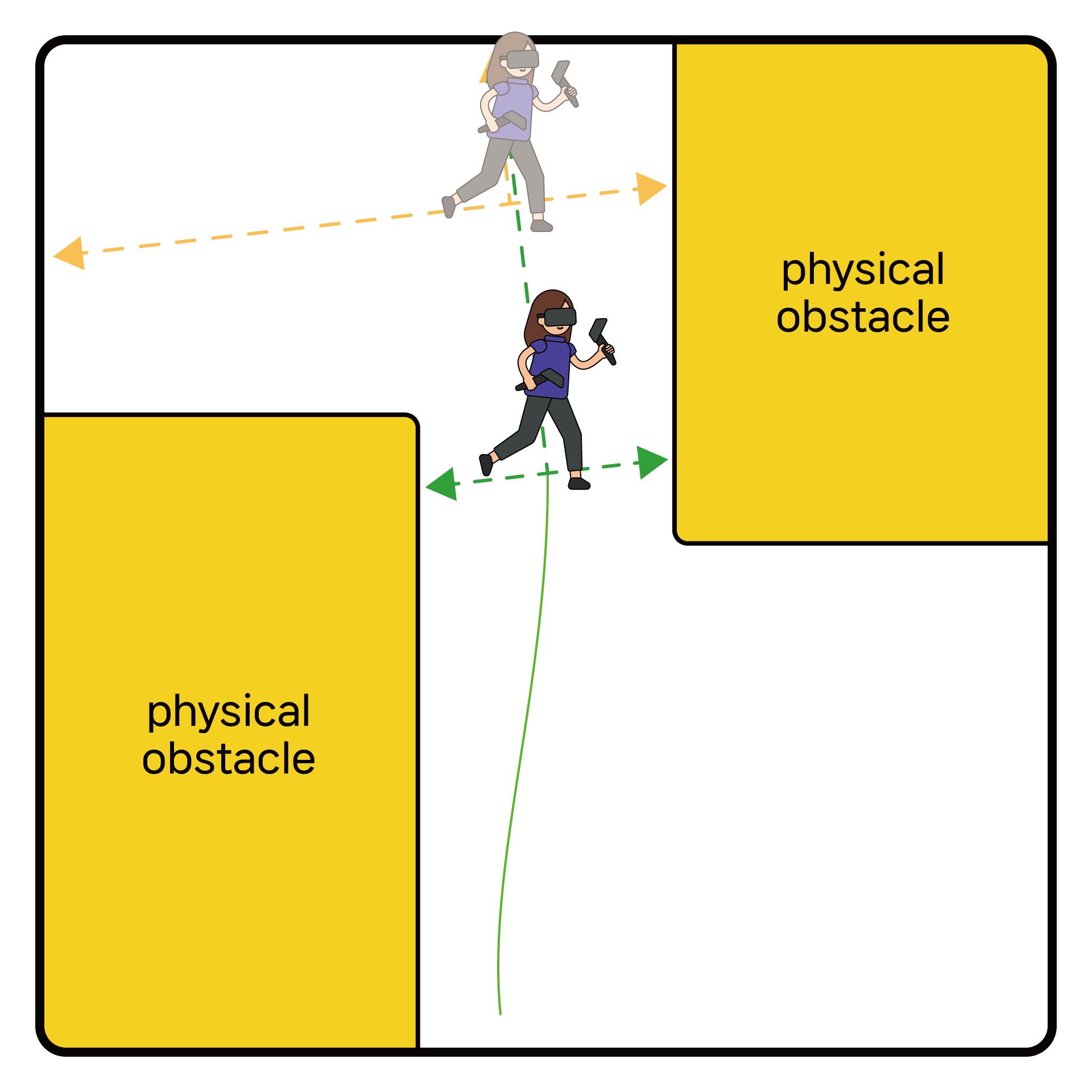}}
        \vspace{-2mm}
        \caption{Demonstration of F-ARC -- one of the RDW methods combined with our proposed mechanism: (a) Our mechanism is empowered by machine learning to predict the user's future position (transparent) in the virtual space. (b) In the given physical and virtual space, the vanilla ARC method only compares the current redirection alignments (pink dotted lines) to redirect users to a direction (pink solid line) that minimizes their misalignment. (c) F-ARC computes the future redirection alignment (yellow dotted lines) based on the predicted future position and then redirects users (green solid line) by considering both current and future redirection alignment.}
        \label{fig:F-ARC}
    \end{teaserfigure}

\maketitle

    \begin{figure*}[t]
        \centering
            \subfigure[\label{fig:PreviousWork01}]
            {\includegraphics[width=0.3\textwidth]{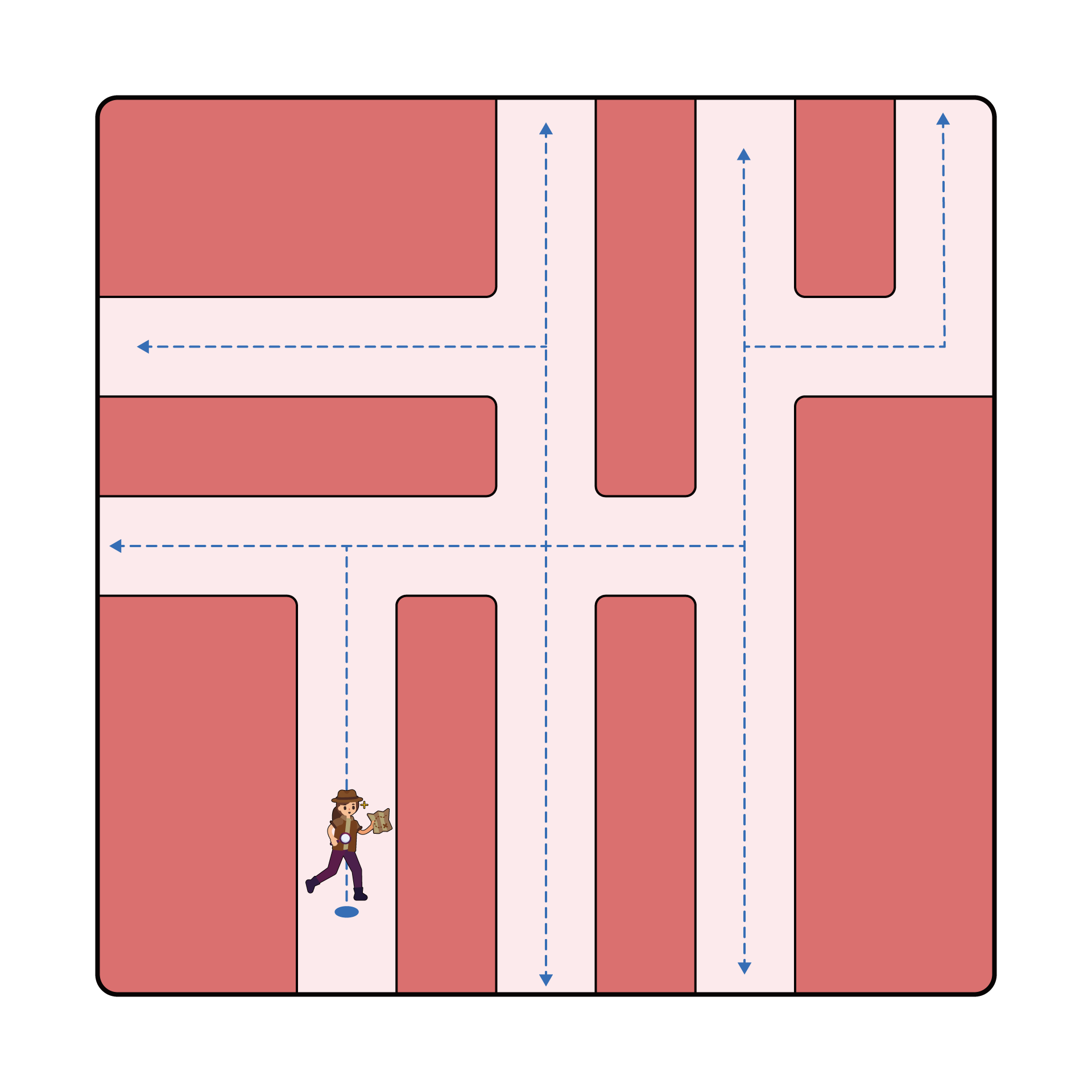}}
            \hspace{0.1cm}
            \subfigure[\label{fig:PreviousWork02}]
            {\includegraphics[width=0.3\textwidth]{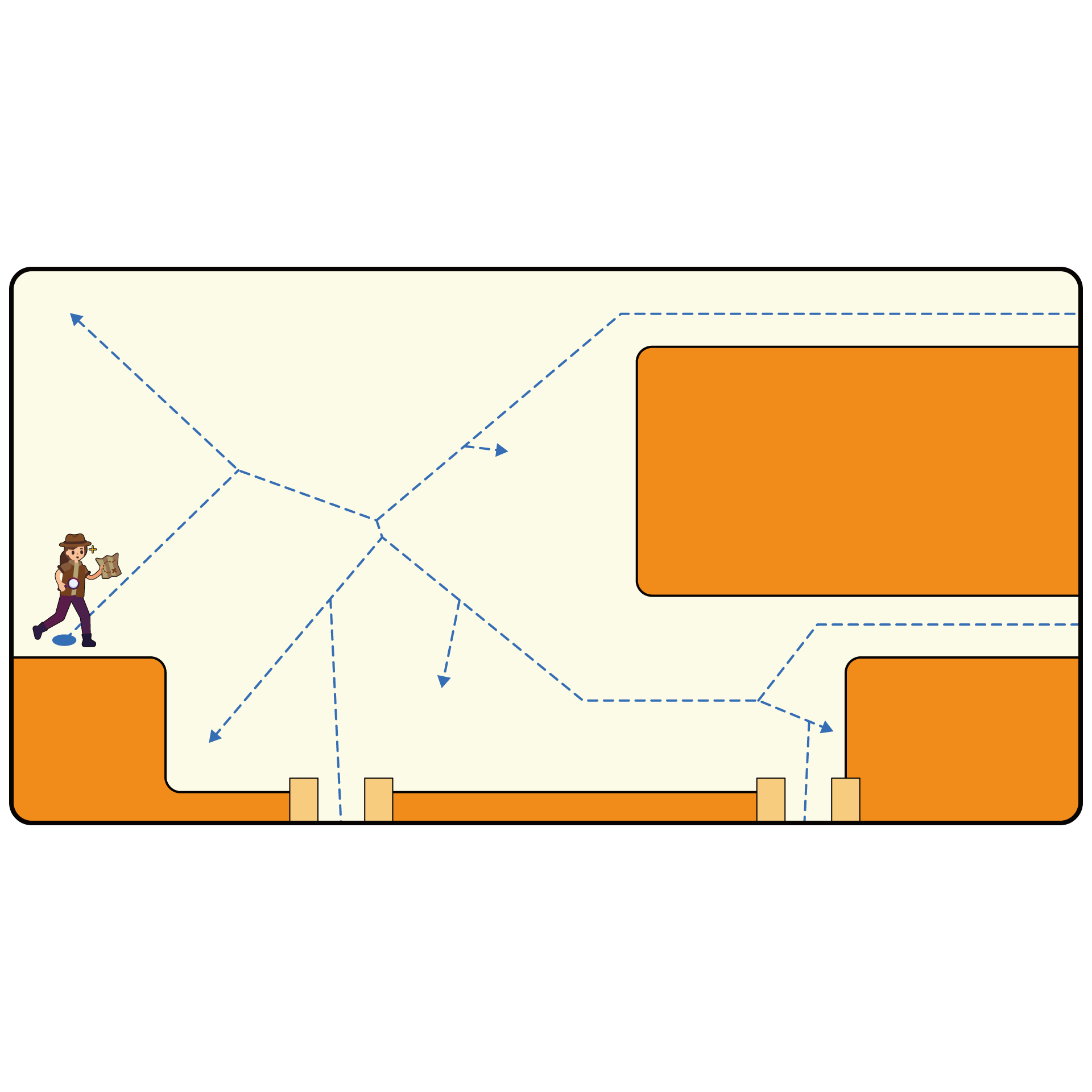}}
            \hspace{0.1cm}
            \subfigure[\label{fig:PreviousWork03}]
            {\includegraphics[width=0.3\textwidth]{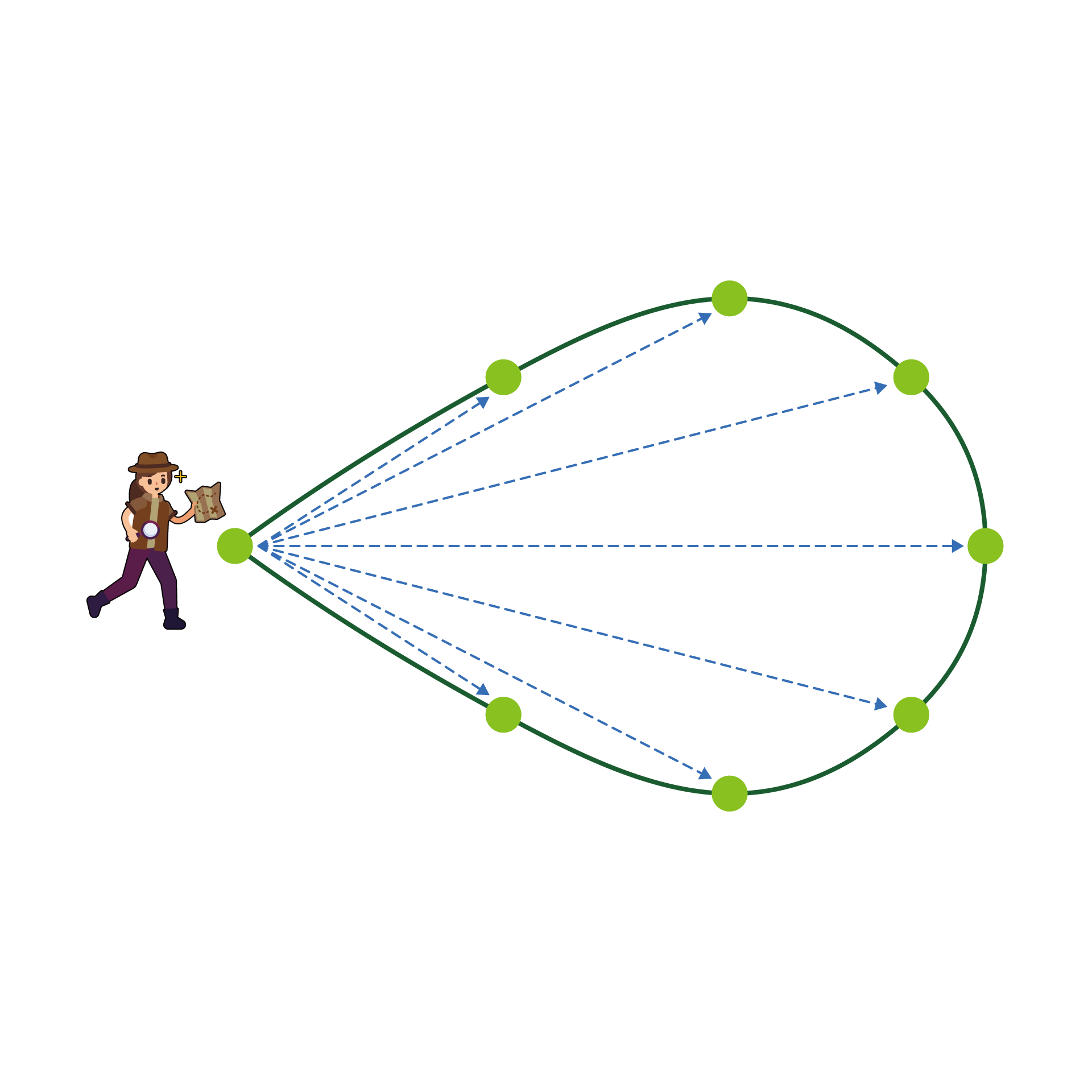}}
        \vspace{-2mm}
        \caption{Limitations of previous RDW methods: (a) Restricting the layout of the virtual space into a maze in order to reduce the number of possible future outcomes. (b) Pre-defining the path graph with virtual paths generated from the Voronoi Diagram, under the assumption that users will always explore along the center line of the paths. (c) Pre-defining the path graph by constructing a discretized virtual path with the assumption that users will advance in the direction he or she is facing. }
        \label{fig:PreviousWork}
        \vspace{-3mm}
    \end{figure*}

\vspace{-2mm}
\section{Introduction}
In virtual reality, locomotion interfaces that allow users to explore virtual environments with a high sense of presence are actively investigated in many studies. Among those interfaces, natural walking is particularly known to provide a higher level of presence to users \cite{usoh1999walking}. However, it is not feasible to rely only on natural walking when exploring infinite virtual space since the corresponding physical space, where users are present, is essentially finite \cite{dong2021tailored,li2021synthesizing,serrano2020imperceptible}. To address this fundamental predicament, many Redirected Walking (RDW) -- a method that mutates the virtual environment or maneuvers the locomotion mapping ratio between the physical and virtual space that the user is engaged -- methods were introduced \cite{Razzaque2001redirected,steinicke2010change,suma2011leveraging}. Previous studies support that users can feel a higher sense of immersion and presence than any other locomotion interfaces when using RDW to explore infinite virtual environments \cite{steinicke2009real,steinicke2009estimation,langbehn2018evaluation}.

Overt technique is a type of redirection technique that instantly guides users inwards to prevent them from crossing over the physical boundary. Among various overt techniques, reset techniques are commonly used in conjunction with subtle techniques. However, reset techniques explicitly ask users to rotate in place, which interferes with the sense of immersion and presence of users in the virtual space. Therefore, predictive RDW methods aim to reduce resets and provide better RDW experiences to users by advising redirection based on the user's future information \cite{suma2012taxonomy, zmuda2013optimizing,nescher2014planning}.

However, the performance of such predictive RDW methods is highly dependent on the prediction accuracy of the future information of the user \cite{nilsson201815}. Moreover, the number of possible future spatial information in an infinitely large virtual space is nearly infinite, which requires mass computational cost, hindering real-time applications \cite{zank2017optimized,azmandian2022adaptive}.

Existing approaches seek to overcome such challenges by making assumptions on the layout of the virtual environment or the walking direction of users. As Figure \ref{fig:PreviousWork} (a) illustrates, early attempts limited the layout of the virtual space to a maze or depended on a pre-defined virtual path graph to reduce the number of possible future walking directions of users \cite{zmuda2013optimizing,nescher2014planning,zank2017optimized,hirt2019prewap}. Following these studies, considerable predictive RDW methods in open virtual space were introduced, but they assumed particular walking direction of users as shown in Figure (b,c) \cite{zank2016you,hirt2019short,hirt2022heuristic,hirt2022short,dong2020dynamic,li2022segmented,azmandian2022adaptive,thomas2022inverse}.
In the meantime, significant progress was made in predicting future locomotion by using spatial or eye-tracking data, but there was no investigation on how the predicted values can be reflected in RDW algorithms \cite{nescher2012analysis,gandrud2016predicting,zank2016eye,bremer2021predicting,stein2022eye,cho2018path,lemic2022short}. 

To overcome the limitations posed by existing methods addressed above, we propose a novel predictive redirection mechanism, \emph{Redirected Walking with Forecasting Future Position} (F-RDW). The backbone of the first step of F-RDW is a recurrent neural network trained with spatial data and eye-tracking data of HMD users, which predicts the future position of users in the virtual space. Then, the predicted values are reflected in existing RDW methods in an applicable manner to enhance their performance while respecting their internal mechanism. We applied F-RDW to four existing RDW methods: MPCRed \cite{nescher2014planning}, S2C \cite{razzaque2005redirected}, TAPF \cite{thomas2019general}, and ARC \cite{williams2021arc}.


Our work has the following contributions:
    \begin{enumerate}
        \item We propose a predictive redirection mechanism that is independent of any constraints on the layout of the virtual space or any assumptions on the user's walking direction.
        \item We suggest a flexible strategy that can predict, maneuver, and reflect the future position information of the user in the virtual space into various existing RDW methods that can enhance their performance.
    \end{enumerate}

\section{Background}
\subsection{Redirection Technique}
Based on the previous works from the last two decades, recent RDW methods are actively suggested by many researchers \cite{sun2018towards,langbehn2018blink,dong2019redirected,williams2021redirected,han2022foldable,hoshikawa2022redirecteddoors,kwon2022infinite,thomas2022inverse}. Thus, survey studies that illustrate the evolution of RDW studies often imply the next objective of RDW methods \cite{suma2012taxonomy, nilsson201815, fan2022redirected, li2022comprehensive}. RDW, which was first designed by Razzaque et al. ~\cite{Razzaque2001redirected}, is a technique that allows users to explore larger virtual space even in small physical space by redirecting their walking. This RDW can be configured with multiple redirection techniques (RET), such as subtle techniques and overt techniques \cite{suma2012taxonomy}. To redirect users, subtle techniques adjust redirection gain, which is a locomotion mapping ratio between the physical and virtual space, in order to keep users from noticing any artificial changes. In contrast, overt techniques prompt apparent notifications to users through their display or sound effects for redirection. Despite their upfront behaviors, the reset technique \cite{williams2007exploring} is one of the most widely used overt techniques in RDW methods which generally asks the users to rotate in place if they reach the boundary of a walkable physical space.

Moreover, users may experience motion sickness or lower immersion if a redirection gain that is higher than the threshold is applied \cite{steinicke2009real, steinicke2009estimation, langbehn2017bending, zhang2018detection, williams2019estimation, bolling2019shrinking, li2021detection, cho2021walking, kim2021adjusting,kim2022effects,you2022strafing}.
Unlike subtle techniques, the reset technique uses a rotation gain that exceeds the threshold, negatively impacting user immersion and sense of presence. Therefore, this is a great motivation for many RDW studies to use alternate subtle techniques as much as possible to minimize the reset occurrences or to design better reset direction \cite{thomas2019general, zhang2022adaptive, zhang2022one, xu2022making}. 

    \begin{figure}[t]
        \centering
            \includegraphics[width=0.4\textwidth]{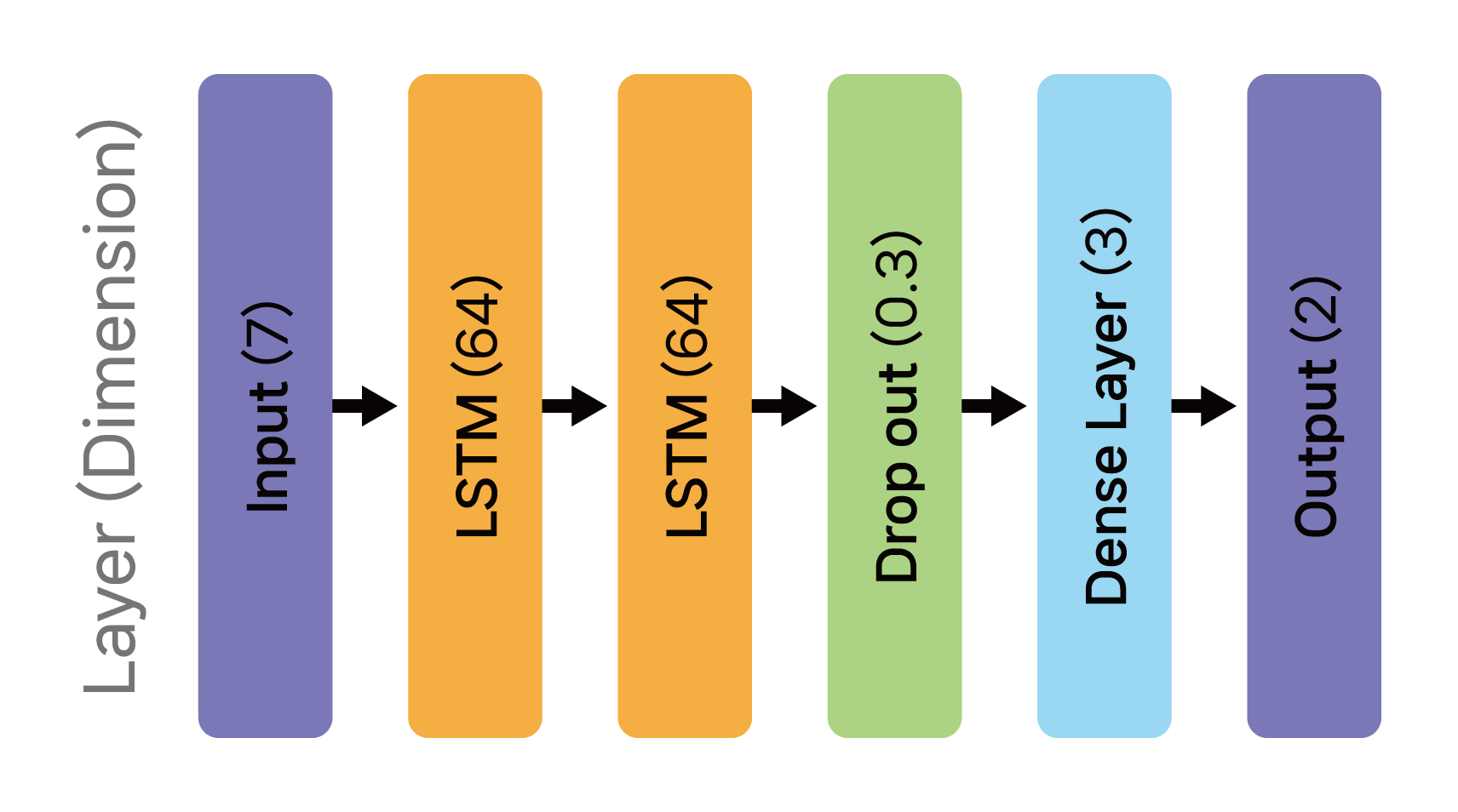}
        \vspace{-2mm}
        \caption{Neural network architecture of the prediction model of F-RDW (F-S2C, F-TAPF, and F-ARC). For F-MPCRed, a softmax function is added after the dense layer in order to output a 1D vector that contains the probabilities that the user will move forward, leftward, and rightward at a given timestamp, respectively.}
        \label{fig:networkarchitecture}
        \vspace{-4mm}
    \end{figure}

\subsection{Predictive Redirection Controller}
Redirection controllers combine subtle and overt techniques which are mainly reactive or possess predictive properties.
Reactive controller (such as Steer-to-Center (S2C) \cite{razzaque2005redirected}, Steer-to-Orbit (S2O) \cite{razzaque2005redirected}, APF-RDW \cite{hoffbauer2018multi}, and ARC \cite{williams2021arc}) is a RET-based algorithm that leads users to a certain point or with a certain pattern.
Predictive controller (such as Fully Optimized Redirected Walking for Constrained Environment (FORCE) \cite{zmuda2013optimizing} or Model Predictive Control Redirection (MPCRed) \cite{nescher2014planning}) is an algorithm that predicts the future travel path of users based on the information of the physical and virtual environment to execute the optimal RET \cite{zank2017optimized,hirt2019prewap, zank2016you,hirt2019short,hirt2022heuristic,hirt2022short}. 
However, previous methods assume a particular layout of the virtual environment or a particular walking direction of users when predicting their future paths (Figure \ref{fig:PreviousWork}).

In the meantime, reinforcement-learning-based approaches for RDW have also been suggested \cite{lee2019real, strauss2020steering, lee2020optimal, shibayama2020reinforcement, chang2021redirection, chen2021reinforcement, wang2022transferable}. Such models can identify and predict the user's path pattern by ingesting past and current spatial information \cite{jeon2022dynamic}. However, there were no attempts to use user-specific data (e.g., gaze direction) in reinforcement learning to make better path predictions in RDW studies. If user-specific data are to be used in reinforcement learning, a separate encoder is needed for better feature extraction \cite{moon2022speeding}, and its influence on the RDW problem should be examined.

    \begin{figure}[t]
        \centering
            \includegraphics[width=0.4\textwidth]{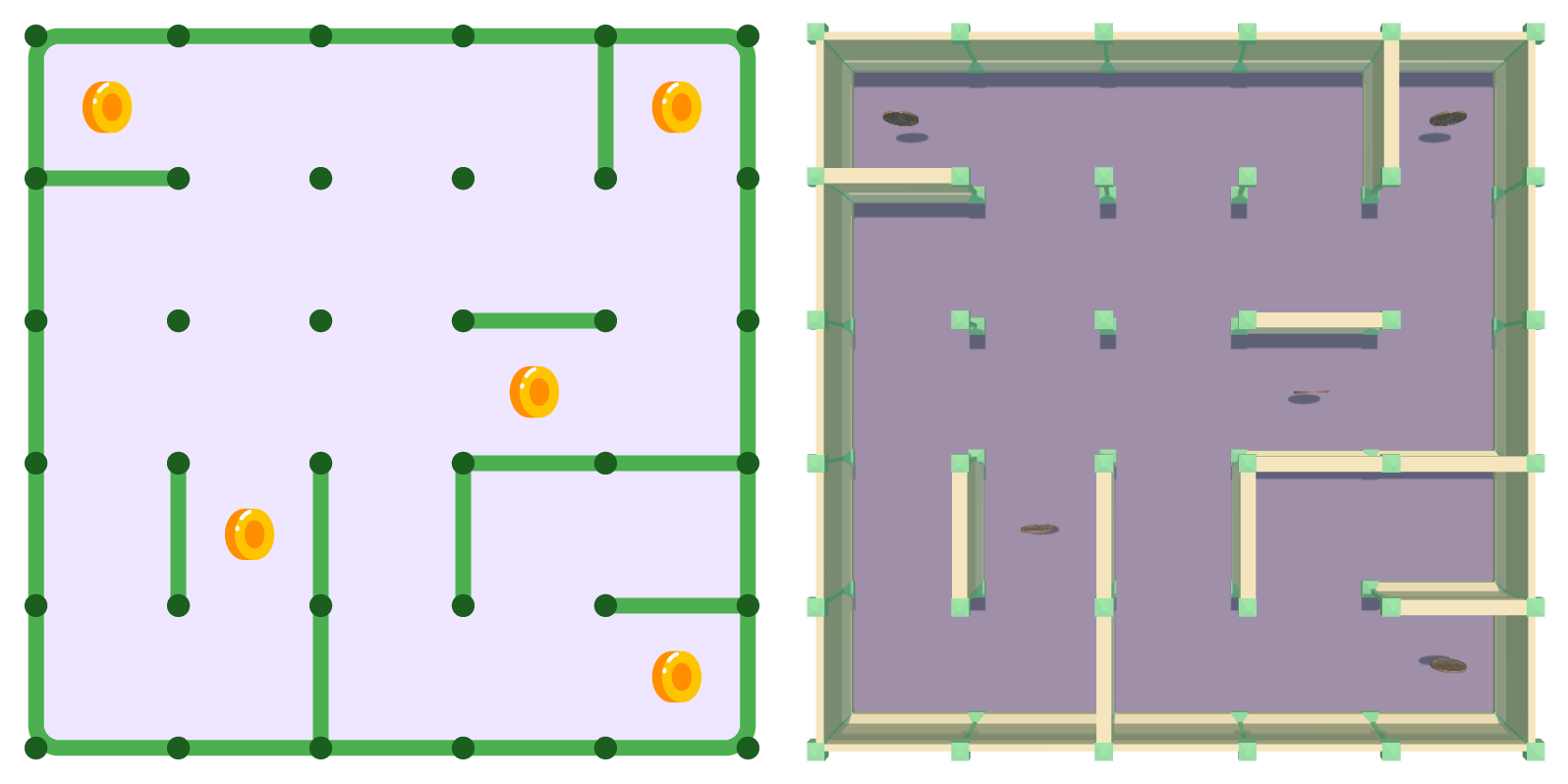}
        \caption{Virtual environment used for simulation tests and generating train and test dataset for the prediction model of F-RDW. For each trial, 10--15 wall objects with 4$m$ width each are randomly generated within a square virtual space with $20m \times 20m$ dimensions.}
        \label{fig:simulationEnv_VE}
    \vspace{-2mm}
    \end{figure}

        \begin{table}[t]
        \centering
        \caption{
        The hyperparameters for the prediction model of F-RDW
        }
        \vspace{0mm}
        {
        \begin{tabular}{lllll}
            \cline{1-5}
            Hyperparameter      & Value &  Hyperparameter      & Value &  \\\cline{1-5}
            Learning Rate & 0.0005 & Sequence Length & 50 & \\
            Epoch & 100 & Hidden Unit & 64 & \\
            BatchSize & 64 & Optimizer & Adam & \\\cline{1-5}
        \end{tabular}
        }
        \label{table:hyperparameters1}
        \vspace{0mm}
    \end{table}

\subsection{Forecasting Future Position}
Acquiring the future position of a user in the virtual space beforehand is a great driving force for efficient path planning for VR locomotion \cite{cho2018path,stein2022eye,bremer2021predicting}. Based on this idea, spatial data or eye-tracking data were ingested by previous machine learning models to predict the user's future position in the virtual space \cite{bremer2021predicting,stein2022eye,lemic2022short,cho2018path}. While some of the studies closely investigated in conjunction with RDW, there has been no strict attempt on how the predicted values can be actually reflected into RDW algorithms and improve their performances.

 \begin{figure}[t]
        \centering
            \subfigure[\label{fig:MPCRed_VE}]
            {\includegraphics[width=0.23\textwidth]{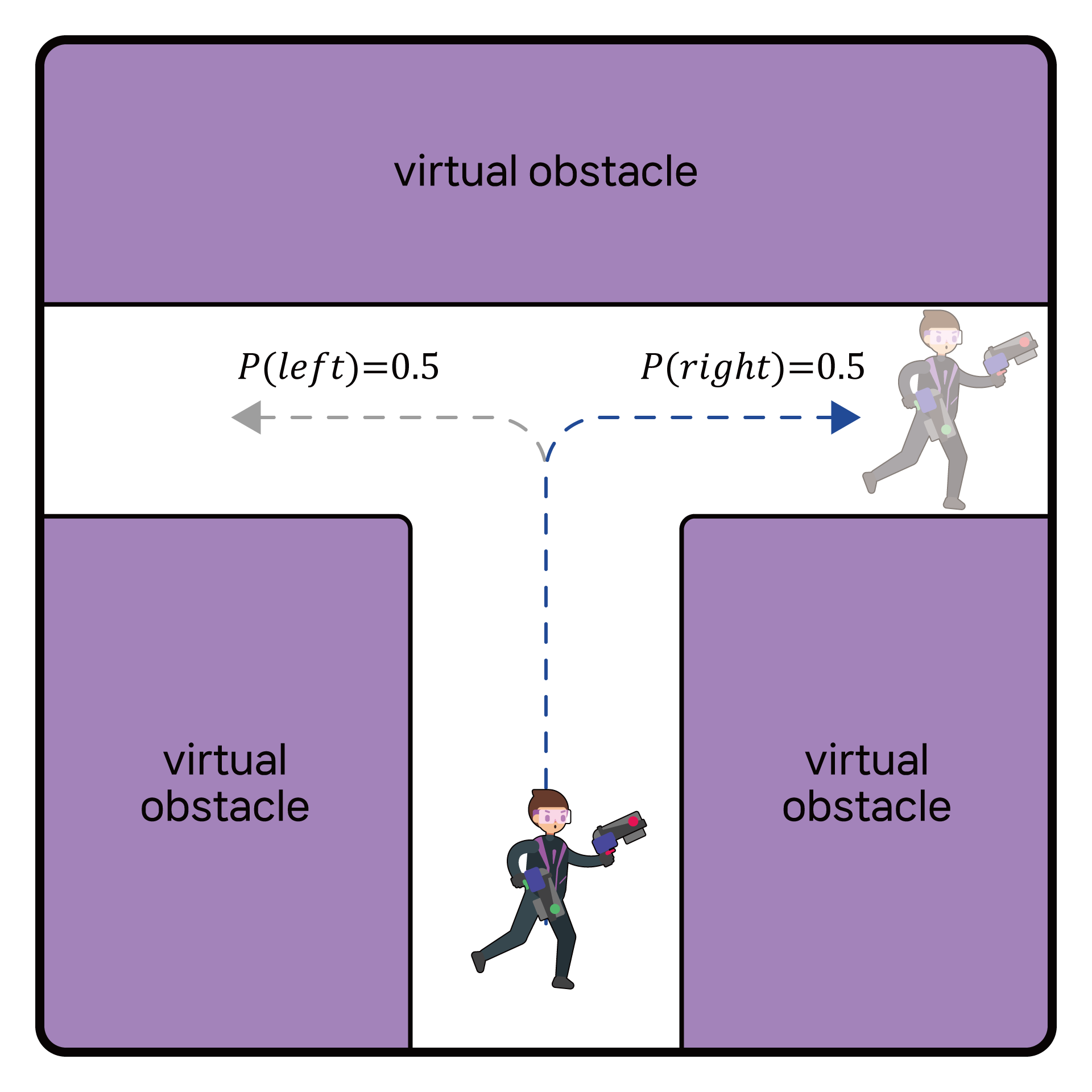}}
            \hspace{0.1cm}
            \subfigure[\label{fig:MPCRed_PE}]
            {\includegraphics[width=0.23\textwidth]{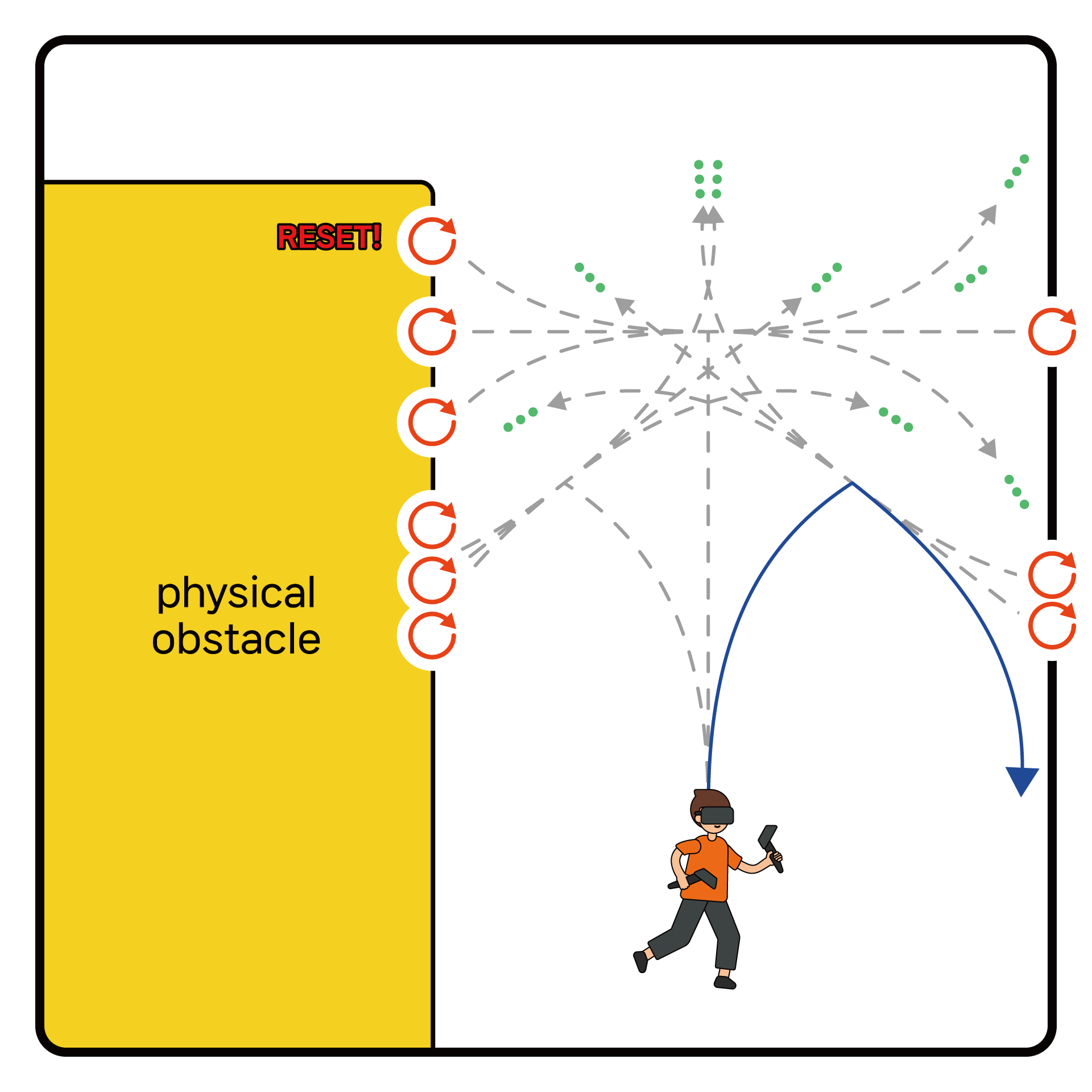}}
        \vspace{-3mm}
        \caption{Limitations of the vanilla MPCRed: (a) The vanilla MPCRed assumes equal possibilities for all possible directions that the user can walk in the virtual space (uniform distribution). (b) In the given physical space, the vanilla MPCRed is likely to select the right curvature under the uniform distribution assumption since selecting the left curvature or no redirection has a high redirection cost due to the high possibility of encountering a reset at an internal obstacle. However, note that redirecting rightwards (blue solid line) may directly encounter a reset since the user will move to the right in the virtual space corner.}
        \label{fig:MPCRed}
        \vspace{-2mm}
    \end{figure}

    \begin{figure}[t]
        \centering
            \subfigure[\label{fig:F-MPCRed_VE}]
            {\includegraphics[width=0.23\textwidth]{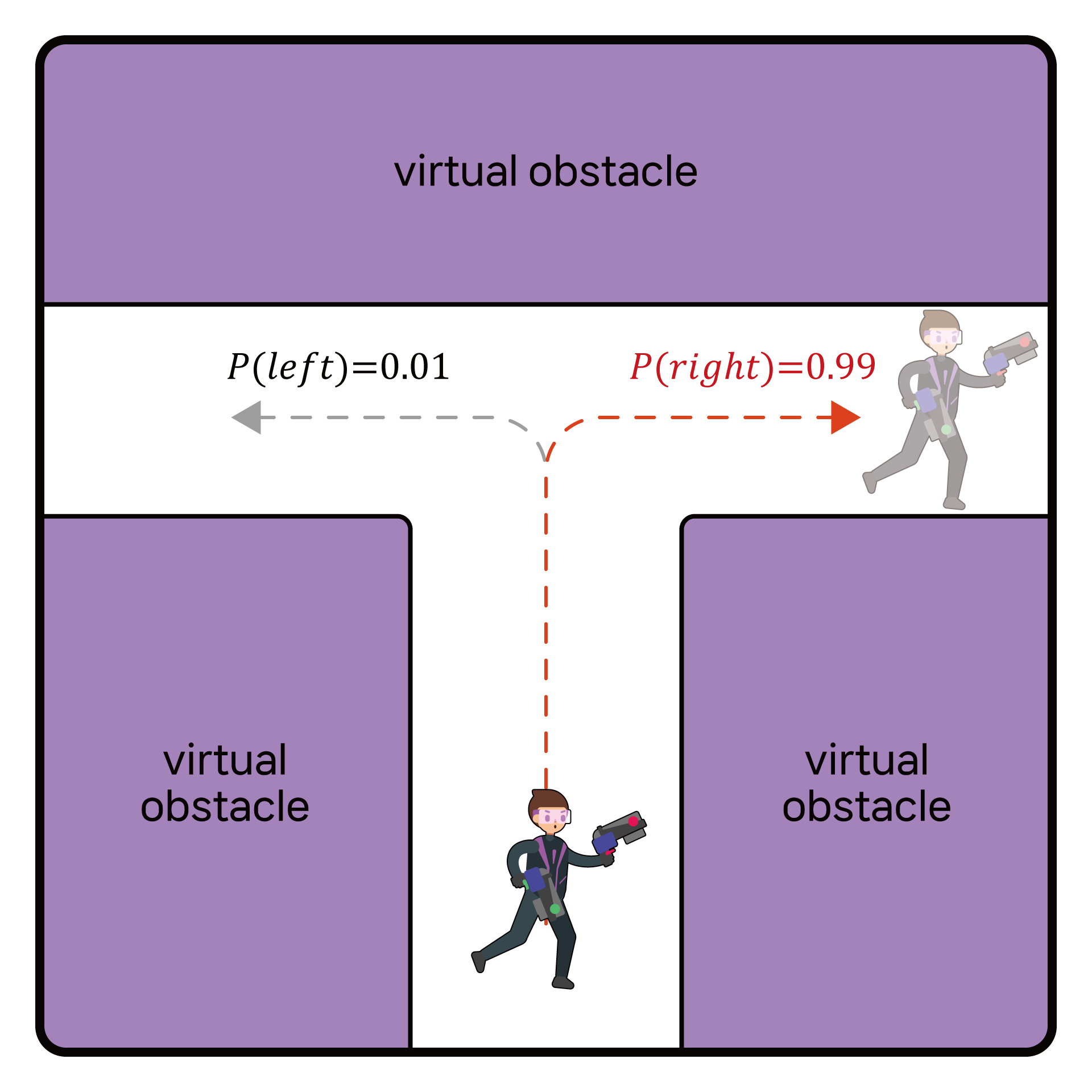}}
            \hspace{0.1cm}
            \subfigure[\label{fig:F-MPCRed_PE}]
            {\includegraphics[width=0.23\textwidth]{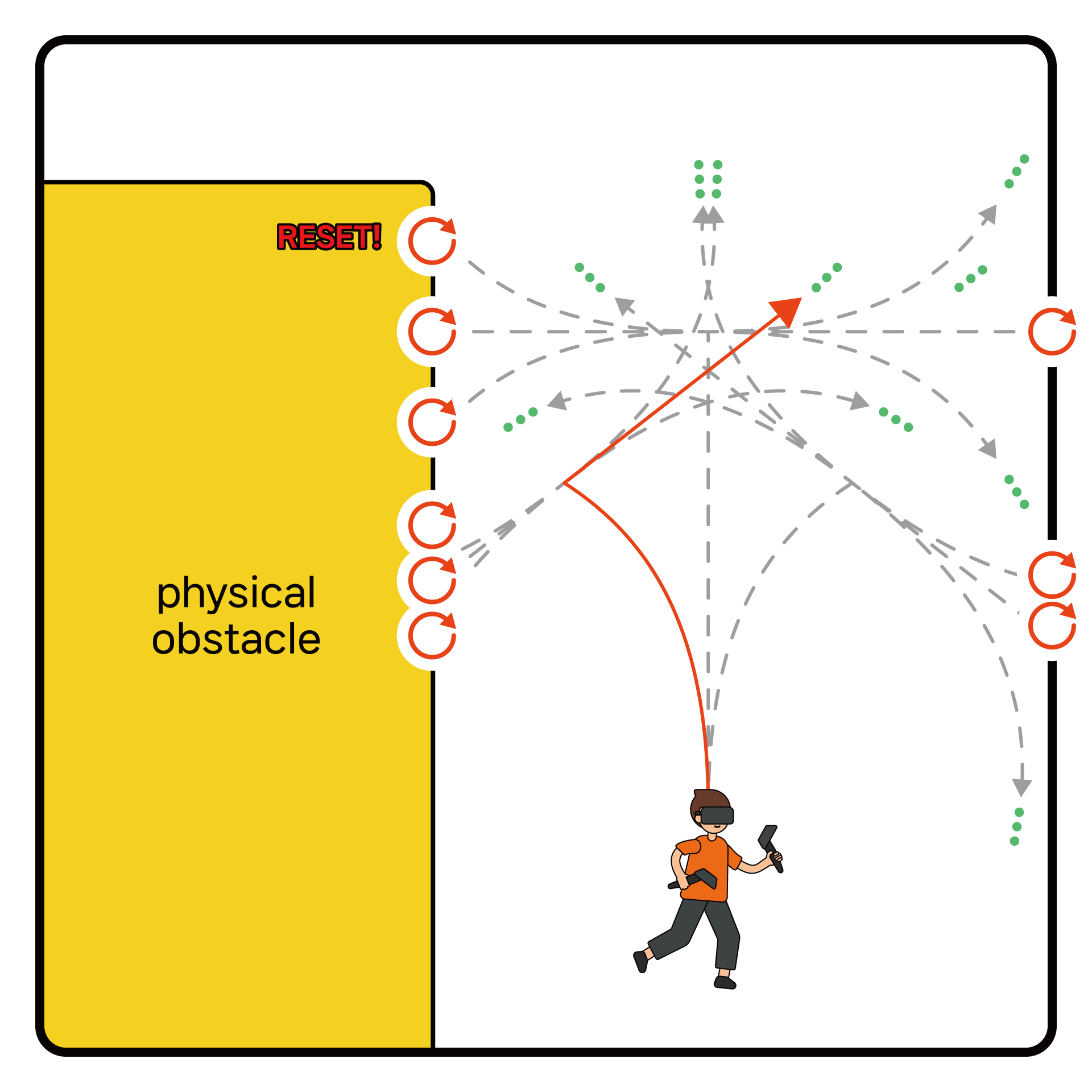}}
        \vspace{-3mm}
        \caption{A winning case of F-MPCRed: (a) F-MPCRed determines the redirection probability per direction by reflecting the future position information acquired from the neural network when encountering a fork in the virtual space. (b) Since our model outputs more polarized per-direction probabilities, the left curvature that was excluded as a candidate in the vanilla MPCRed is now reconsidered to be selected (red solid line).
        This eventually encourages F-MPCRed to redirect users more distant from the physical boundary than the vanilla MPCRed.
        }
        \label{fig:F-MPCRed}
        \vspace{-2mm}
    \end{figure}

\section{Method}         
\subsection{Predicting Future Position}
Our proposed mechanism -- F-RDW -- predicts the future position information of a user in the virtual space without depending on any assumptions about the virtual environment's layout or the user's walking behavior. Based on Bremer et al. [\citeyear{bremer2021predicting}]'s model that is capable of predicting future locomotion,
we designed an LSTM-based recurrent neural network and trained our model with spatial data and eye-tracking data of users (Figure \ref{fig:networkarchitecture}). The input of our model consists of 2D gaze direction, 1D body orientation (up-axis), 2D neck orientation, and 2D velocity vector. The output of our model varies depending on which RDW method it is combined with. For instance, our model may output a 2D vector that represents the future position of a user in the virtual space or output a 1D vector that denotes the walking probability towards each cardinal direction (except backward).

In order to collect the training data for our network, we recruited 10 participants to engage in a VR experience. The participants were asked to walk around and collect 3--5 virtual targets (coins) which were randomly generated in a virtual environment (Figure \ref{fig:simulationEnv_VE}) that was mapped into an empty $4m \times 4m$ squared physical space.
Note that we only applied the Reset-to-Center (R2C) \cite{thomas2019general} technique to redirect the users when they were exploring the virtual space with RDW.
Under such an environment, we collected time-series data of 120,000 frames shot with an average of 60Hz. We trained our model with Intel i7-11700 (16GB) and RTX 3070 for 20 minutes by using the hyperparameters described in Table \ref{table:hyperparameters1} and validated our mechanism with K-fold cross-validation ($K=5$). According to the results of our simulation test, our model can correctly predict the future walking direction (in cardinal directions except backward) of a user with 77\% accuracy and predict the future position of a user with a mean displacement error (MDE) of $M=0.45m$ with $SD=0.35m$.

    \begin{algorithm}[t]
        \caption{F-MPCRed}\label{alg:F-MPCRed}
        \KwIn
        {
            \\
            \quad \,\, $u^v_c$: user's current 2D virtual position, \\
            \quad \,\, $\gamma$: \textbf{user's future movement direction probability
            }
        }
        \SetKwFunction{FMain}{$F\textbf{-}MPCRed$}
        \SetKwProg{Fn}{Function}{:}{}
        \Fn{\FMain{$u^v_c, \gamma$}}
        {
            $\alpha = 0.8, k = 4$; \\
            $bestCost \gets \infty$; 
            \\
            \For{$a \in All Redirection Action$}
            {
                $cost_{c} \gets 0$;
                \\
                \If{$cost(a) < bestCost$}
                {
                    \For{$s \in GetVirtualPathSegments(u^v_c)$} 
                        { 
                            $[u^v_{next}, cost_{stage}] \gets APPLY(a, u^v_c, s)$; \\
                            $cost_{c} \gets cost_{c} + $ \textbf{$\gamma$} $ * cost_{stage}$; 
                            \\
                            \If{$cost_{c} \geq bestCost$}
                            {
                                \textbf{break}
                            }

                            \If{$k > 0$}
                            {
                                $[a_{next}, cost_{next}] \gets F\textbf{-}MPCRed(u^v_{next}, k-1, \gamma)$; \\
                                $cost_{c} \gets cost_{c} + \alpha * $ \textbf{$\gamma$} $ * cost_{next}$; \\
                            }
                        }
                        \If{$cost_{c} < bestCost$}
                        {
                            $bestCost \gets cost_{c}$; \\
                            $bestAction \gets a$; \\
                        }                
                }

            }
            $RedirectUser(bestAction)$; \\
        }
        \textbf{End Function}
    \end{algorithm}

  \begin{figure*}[t]
        \centering
            \subfigure[\label{fig:S2C_VE}]
            {\includegraphics[width=0.35\textwidth]{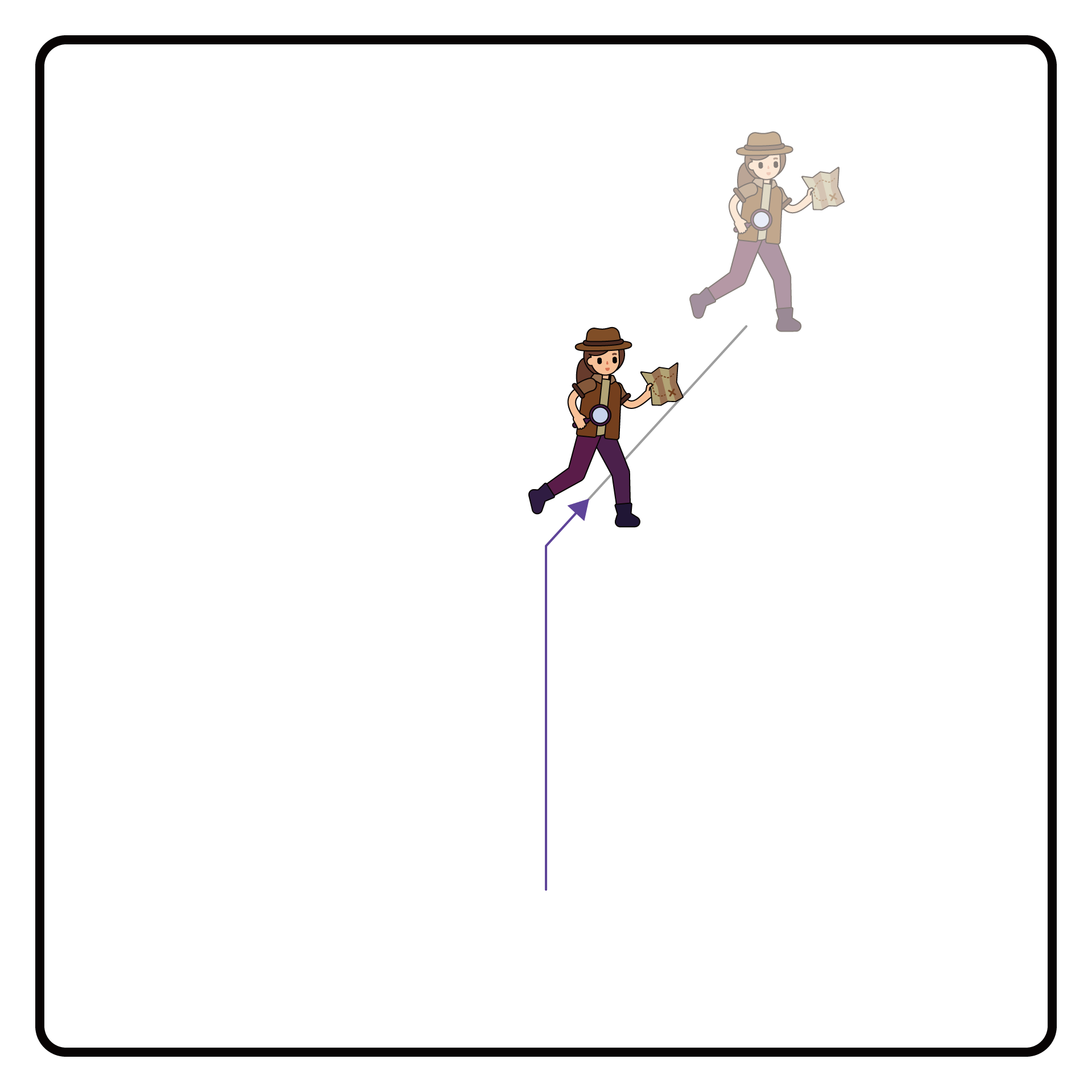}}
            \hspace{0.1cm}
            \subfigure[\label{fig:S2C_PE}]
            {\includegraphics[width=0.3\textwidth]{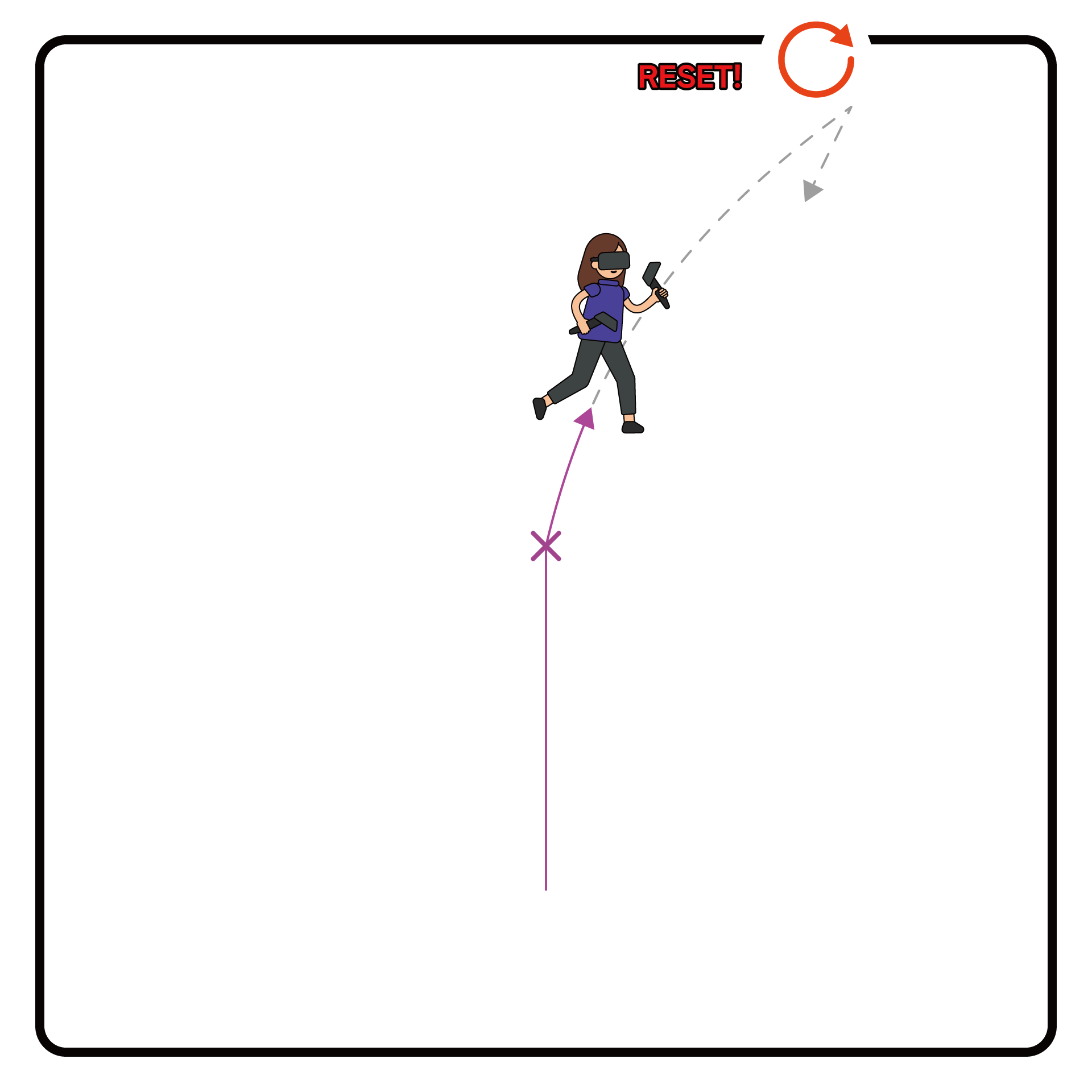}}
            \hspace{0.1cm}
            \subfigure[\label{fig:F-S2C_PE}]
            {\includegraphics[width=0.3\textwidth]{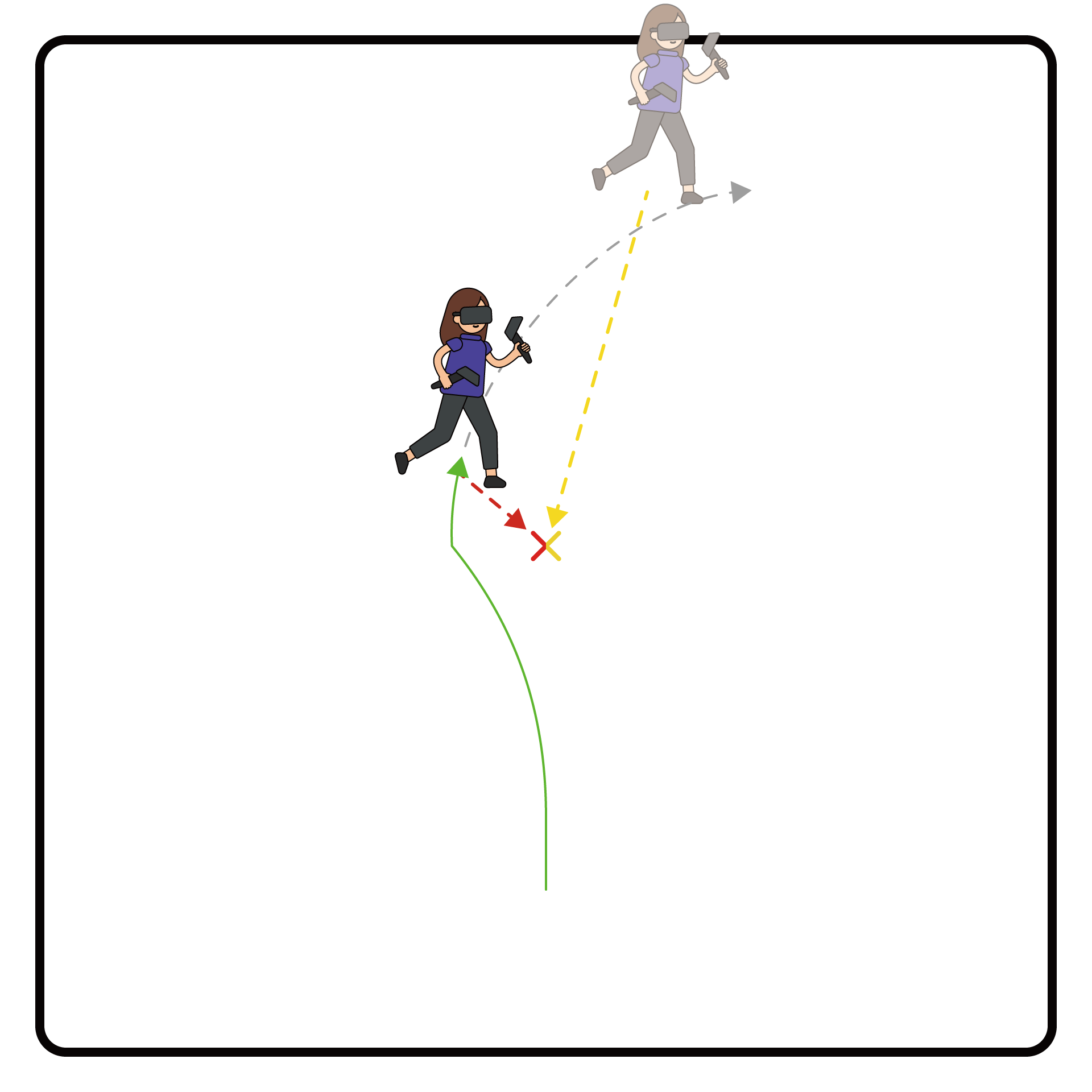}}
        \vspace{-3mm}
        \caption{Mechanism of F-S2C (S2C combined with our proposed mechanism): (a) F-RDW predicts the virtual future position (transparent) of a user at $F_t=1$ (timestamp of the forecasted point) by using its prediction model. (b) The vanilla S2C algorithm greedily redirects a user towards the redirection target that is located in the center point. (c) F-S2C overlays the user's future position in the virtual space on the physical space (transparent) and redirects the user (green solid line) by considering both directions; direction from the current and future position to the center point, respectively (red and yellow dotted lines). This ultimately makes F-S2C possibly redirect a user longer without encountering a reset than the original S2C method.}
        \label{fig:F-S2C}
    \vspace{-3mm}
    \end{figure*}

    \begin{figure*}
        \centering
        \subfigure[\label{fig:TAPF_VE}]
            {\includegraphics[width=0.35\textwidth]{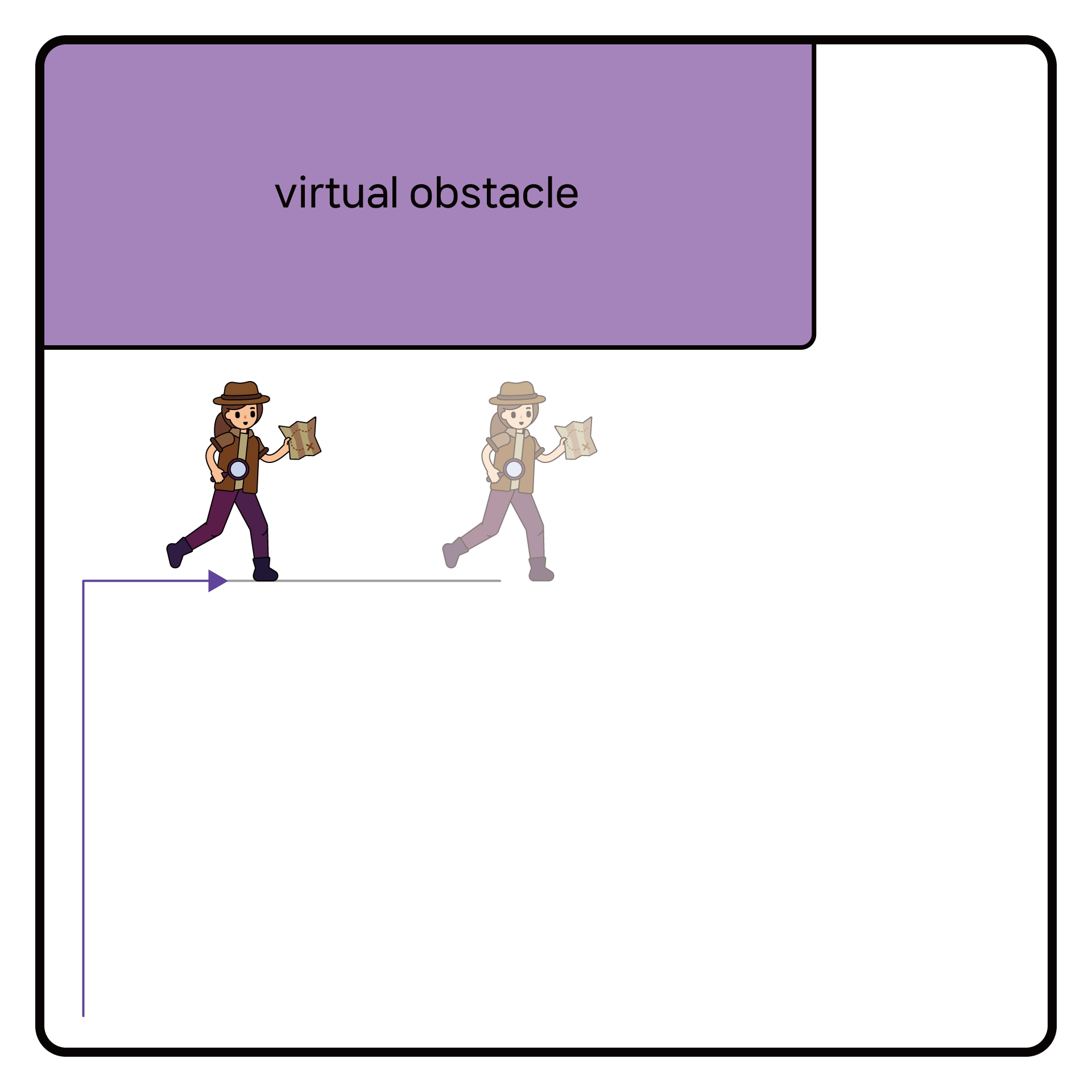}}
            \hspace{0.1cm}
        \subfigure[\label{fig:TAPF_PE}]
            {\includegraphics[width=0.3\textwidth]{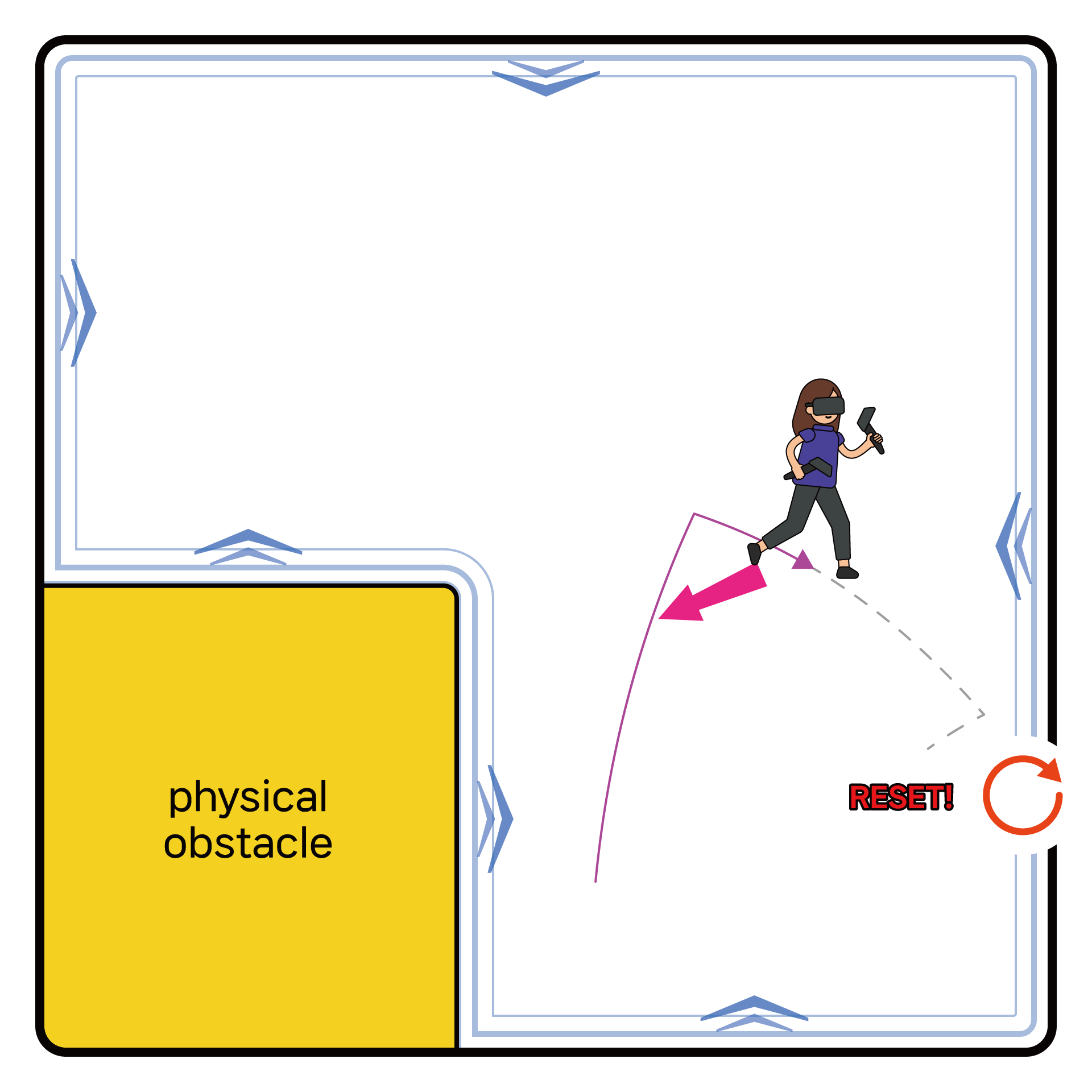}}
            \hspace{0.1cm}
        \subfigure[\label{fig:F-TAPF_PE}] 
            {\includegraphics[width=0.3\textwidth]{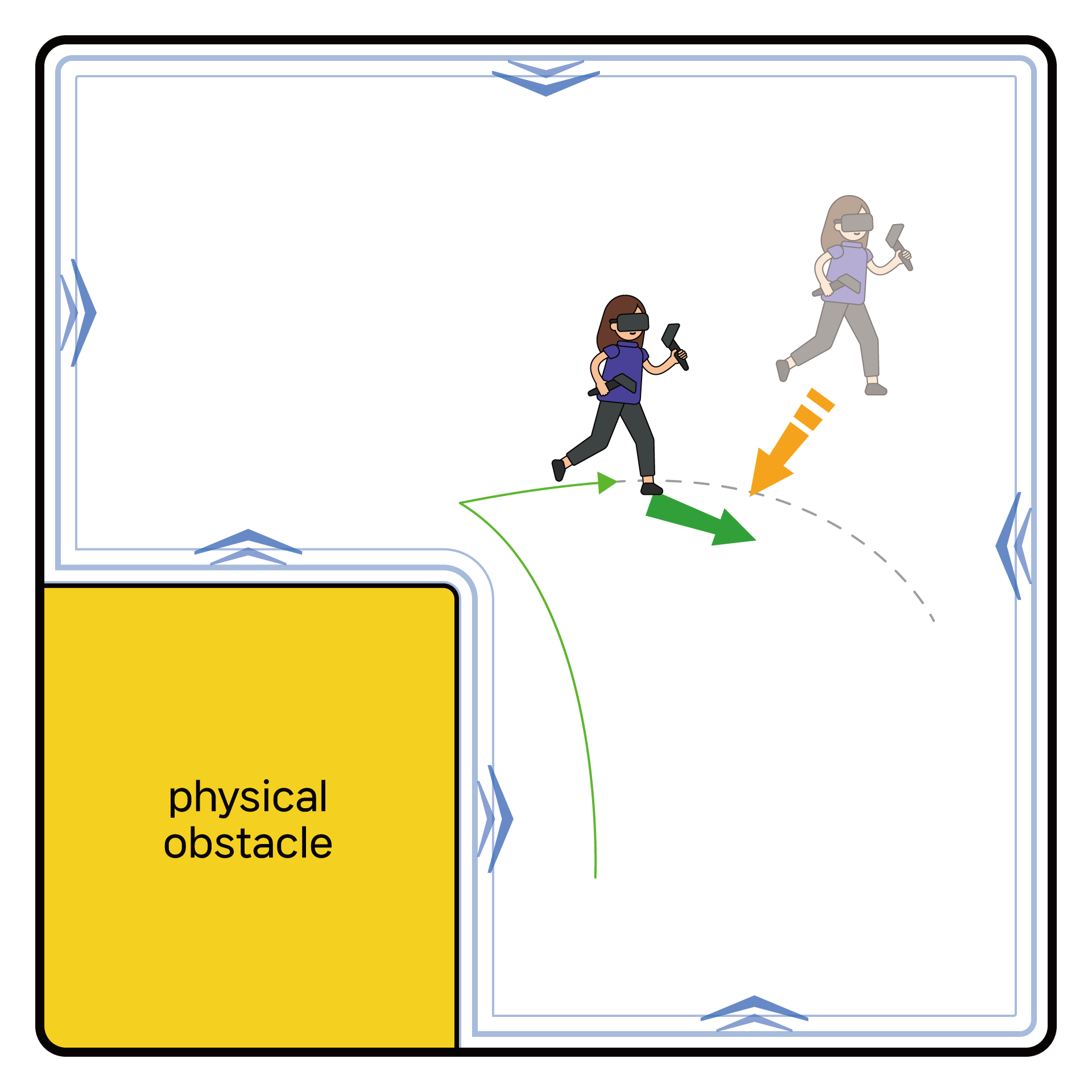}}
        \vspace{-3mm}
        \caption{
        Mechanism of F-TAPF (TAPF combined with our proposed mechanism):
        (a) According to the future position of the user after 1 second (transparent) forecasted by the prediction model of F-RDW, the user is likely to walk in the right direction in the virtual space. In this case, it is ideal for the redirection algorithm to secure walkable space on the right side of the user in the physical space beforehand for better path planning. (b) In the given physical space, the original TAPF algorithm redirects the user in the current resultant direction (pink arrow). (c) In contrast, F-TAPF first determines the net force of two forces -- the resultant force exerted at the future position (orange arrow) and the resultant force exerted at the current position (green arrow). Then, the algorithm redirects the user in the direction of their net force (green line).}
        \label{fig:F-TAPF}
    \vspace{-3mm}
    \end{figure*}

\subsection{Applying Future Position}
One of the strong points of F-RDW is that its predicted future information $f$ can be easily incorporated with any RDW method without altering each internal mechanism. In this work, we demonstrated how F-RDW could be combined with well-known RDW methods, including MPCRed, S2C, TAPF, and ARC, without changing any internal parameters used in each RDW method. We also empirically discovered and applied the optimal $\mu$ values -- the weight between the current and future information used when making predictions for redirection -- for each F-RDW method, which we will discuss in detail in Section 5.1. In the following section, we provide a detailed illustration of how our mechanism fuses with each RDW method.

    \begin{algorithm}[t]
        \caption{F-S2C}\label{alg:F-S2C}    
        \KwIn
        {
            \\
            \quad \,\, $u^v_c$: user's current 2D virtual position, \\  
            \quad \,\, $u^p_c$: user's current 2D physical position,\\ 
            \quad \,\, $u^v_f$: \textbf{user's future 2D virtual position\\ 
            }

        }
        \SetKwFunction{FMain}{$F\textbf{-}S2C$}
        \SetKwProg{Fn}{Function}{:}{}
        \Fn{\FMain{$u^v_c, u^p_c, u^v_f$}}
        {    
            $d_c \gets ComputeVectorToCenterPoint(u^p_c)$; \\
            $u^p_f \gets u^v_f$;\\
            $d_f \gets ComputeVectorToCenterPoint(u^p_f)$; \\

            $r \gets d_f * \mu + d_c * (1 - \mu)$;\\

            $RedirectToTarget(r)$;
        }
        \textbf{End Function} 
    \end{algorithm}

    \begin{algorithm}[t]
        \caption{F-TAPF}\label{alg:F-TAPF}    
        \KwIn
        {
            \\
            \quad \,\, $u^v_c$: user's current 2D virtual position, \\  
            \quad \,\, $u^p_c$: user's current 2D physical position,\\ 
            \quad \,\, $u^v_f$: \textbf{user's future 2D virtual position
            }

        }
        \SetKwFunction{FMain}{$F\textbf{-}TAPF$}
        \SetKwProg{Fn}{Function}{:}{}
        \Fn{\FMain{$u^v_c, u^p_c, u^v_f$}}
        {
            $force_{c} \gets ComputeTAPF(u^p_c)$; \\
            $u^p_f \gets u^v_f$;\\
            $force_{f} \gets ComputeTAPF(u^p_f)$; \\

            $force_{total} \gets force_{f} * \mu + force_{c} * (1 - \mu)$; \\

            $g_t, g_c, g_r \gets CalculRedirGains(force_{total})$; \\
            
            
            $RedirectUsers(g_t, g_c, g_r)$; \\        
        }
        \textbf{End Function} 
    \end{algorithm}

    \begin{algorithm}[t]
        \caption{F-ARC}\label{alg:F-ARC}    
        \KwIn
        {
            \\
            \quad \,\, $u^v_c$: user's current 2D virtual position, \\
            \quad \,\, $o^v_c$: user's current 1D virtual orientation, \\  
            \quad \,\, $u^p_c$: user's current 2D physical position,\\
            \quad \,\, $o^p_c$: user's current 1D physical orientation, \\  
            \quad \,\, $u^v_f$: \textbf{user's future 2D virtual position
            }       

        }
        \SetKwFunction{FMain}{$F\textbf{-}ARC$}
        \SetKwProg{Fn}{Function}{:}{}
        \Fn{\FMain{$u^v_c, o^v_c, u^p_c, o^p_c, u^v_f$}}
        {
            
            ${ml}_c \gets ComputeMisalign(u^v_c, o^v_c, u^p_c, o^p_c)$; \\
            $g_t^c, g_c^c \gets CalculRedirGains({ml}_c)$; \\

            $o^p_f, o^v_f \gets u^v_c - u^v_f$; \\
            $u^p_f \gets u^v_f$; \\
            ${ml}_f \gets ComputeMisalign(u^v_f, o^v_f, u^p_f, o^p_f)$; \\
            $g_t^f, g_c^f \gets CalculRedirGains({ml}_f)$; \\
            $g_t = g_t^f * \mu + g_t^c * (1-\mu)$; \\
            $g_c = g_c^f * \mu + g_c^c * (1-\mu)$; \\

            \eIf{the user rotates toward $u^p_f$}{
                \eIf{${ml}_c > {ml}_f$}{
                    $g_r \gets Min$;
                }{
                    $g_r \gets Max$;
                }
            }{
                \eIf{${ml}_c > {ml}_f$}{
                    $g_r \gets Max$;
                }{
                    $g_r \gets Min$;
                }
            }

            $RedirectUsers(g_t, g_c, g_r)$; \\  
        }
        \textbf{End Function} 
    \end{algorithm}

\subsubsection{F-MPCRed}
The MPCRed algorithm suggested by Nescher et al. [\citeyear{nescher2014planning}] first constructs an event tree for possible scenarios when redirection techniques are continuously applied when a current and terminal location is given. Then, the algorithm selects the optimal RET sequence from those candidates. To implement such behavior, MPCRed conducts a depth-first search for each RET sequence candidate at every path segment step to compute the redirection cost. Then, each redirection cost is multiplied by the probability of its corresponding path being selected; note that the probability has a uniform distribution pre-determined by the pre-defined virtual path graph in this case. Finally, the RET sequence with the minimum product value is selected (Figure \ref{fig:MPCRed}).

On the other hand, F-MPCRed (Algorithm \ref{alg:F-MPCRed}) does not depend on pre-defined or uniform probability per future walking direction. Instead, it imports much more polarized and rational per-direction probabilities acquired by our prediction model, as shown in Figure \ref{fig:F-MPCRed} (a). Hence, more polarized per-direction probabilities are multiplied by the redirection cost computed in Figure \ref{fig:F-MPCRed} (b). This process encourages F-MPCRed to select a RET sequence with fewer reset occurrences than MPCRed due to a better reflectance of the future information and a better coverage of the possible RET sequences that MPCRed would have excluded its consideration. Note that we used the same parameters, such as recursion depth $k$ and decay factor $\alpha$, defined in the original MPCRed algorithm when simulating the F-MPCRed algorithm.

\subsubsection{F-S2C}
The S2C algorithm introduced by Razzaque et al. [\citeyear{razzaque2005redirected}] assumes that the center of the physical space is the safest location from encountering any obstacles or the physical boundary. Thus, the center point is defined as the redirection target $r$, and users are always redirected towards $r$ as shown in Figure \ref{fig:F-S2C} (b).

F-S2C overlays the predicted future virtual position on the physical space (Figure \ref{fig:F-S2C} (c)) and determines redirection by computing the resultant vector of two vectors: a vector from the predicted physical position to the center point and a vector from the current physical position to the center point (Algorithm \ref{alg:F-S2C}). Then, F-S2C applies a curvature gain with respect to that direction and leads the user to be close to the physical center point. Ultimately, the user can secure more physical space in the direction he/she will explore (Figure \ref{fig:F-S2C} (c)).

\begin{table*}[]
    \centering
    \caption{Statistical significance analysis by standard score $z$, standardized score $t$ and probability value $p$ in \textbf{E1-E4} of our simulation tests.}
    \vspace{-2mm}
\centering
\begin{tabular}{ccllll}
\hline
\multicolumn{2}{c}{\multirow{2}{*}{Experiments}} & \multicolumn{4}{c}{RDW methods}                                                                                   \\ \cline{3-6} 
\multicolumn{2}{c}{} & \multicolumn{1}{c}{MPCRed vs F-MPCRed} & \multicolumn{1}{c}{S2C vs F-S2C} & \multicolumn{1}{c}{TAPF vs F-TAPF} & \multicolumn{1}{c}{ARC vs F-ARC} \\ \hline
\multirow{2}{*}{E1}           & \#Reset          & $t(98) = 3.924, p < .001$  & $z(98) = 2.869, p = .004$  & $z(98) = 2.107, p = .017$  & $z(98) = 1.883, p = .03$   \\
                              & MDbR             & $t(98) = -3.163, p = .001$ & $z(98) = -2.577, p = .009$ & $z(98) = -1.932, p = .026$ & $z(98) = -1.911, p = .028$ \\
\multirow{2}{*}{E2}           & \#Reset          & $z(98) = 0.135, p = .446$  & $z(98) = 1.659, p = .097$  & $z(98) = 1.266, p = .205$  & $t(98) = -1.152, p = .25$  \\
                              & MDbR             & $z(98) = 0.648, p = .258$  & $z(98) = -0.809, p = .418$ & $z(98) = -0.05, p = .96$   & $z(98) = 0.848, p = .4$    \\
\multirow{2}{*}{E3}           & \#Reset          & $z(98) = 2.255, p = .012$  & $t(98) = -1.367, p = .173$ & $t(98) = 5.568, p < .001$  & $t(98) = 3.823, p < .001$  \\
                              & MDbR             & $z(98) = -1.98, p = .023$  & $z(98) = 0.875, p = .381$  & $z(98) = -5.48, p < .001$  & $z(98) = -3.242, p = .001$ \\
\multirow{2}{*}{E4}           & \#Reset          & $z(98) = 1.84, p = .032$   & $z(98) = 1.548, p = .121$  & $z(98) = 6.547, p < .001$  & $z(98) = 4.109, p < .001$  \\
                              & MDbR             & $t(98) = -2.52, p = .012$  & $z(98) = -0.547, p = .584$ & $z(98) = -5.911, p < .001$ & $z(98) = -4.04, p < .001$  \\ \hline
\end{tabular}
    \label{table:SimulationTestTable}
    \vspace{-2mm}
\end{table*}

\subsubsection{F-TAPF}
The TAPF algorithm proposed by Thomas et al. [\citeyear{thomas2019general}] is driven by the assumption that repulsive forces are exerted from physical obstacles or the physical boundary and redirects users according to the resultant repulsive force (function $ComputeTAPF$ in Algorithm \ref{alg:F-TAPF}). In Figure \ref{fig:F-TAPF} (b), the original TAPF algorithm computes the resultant force to be right-oriented. However, since the user is expected to move rightwards in the virtual space (Figure \ref{fig:F-TAPF} (a)), using a right-oriented curvature in the current physical space increases the likelihood of a reset (Figure \ref{fig:F-TAPF} (b)).

On the other hand, F-TAPF overlays the predicted future virtual position on the physical space (Figure \ref{fig:F-TAPF} (c)) and redirects the user in the direction of the net force; the net force is a weighted sum of the resultant force at the overlayed physical point and the resultant force at the current user position (Algorithm \ref{alg:F-TAPF}). This encourages the user to be distant from physical obstacles and the boundary even in the future. Consequently, F-TAPF secures more space than the original TAPF in the right direction in Figure \ref{fig:F-TAPF}, which is identical to the direction that the user is expected to move.

    \begin{figure}[t]
        \centering
        \vspace{-3mm}
            \subfigure[\label{fig:simulationEnv_PE1}]
            {\includegraphics[width=0.2\textwidth]{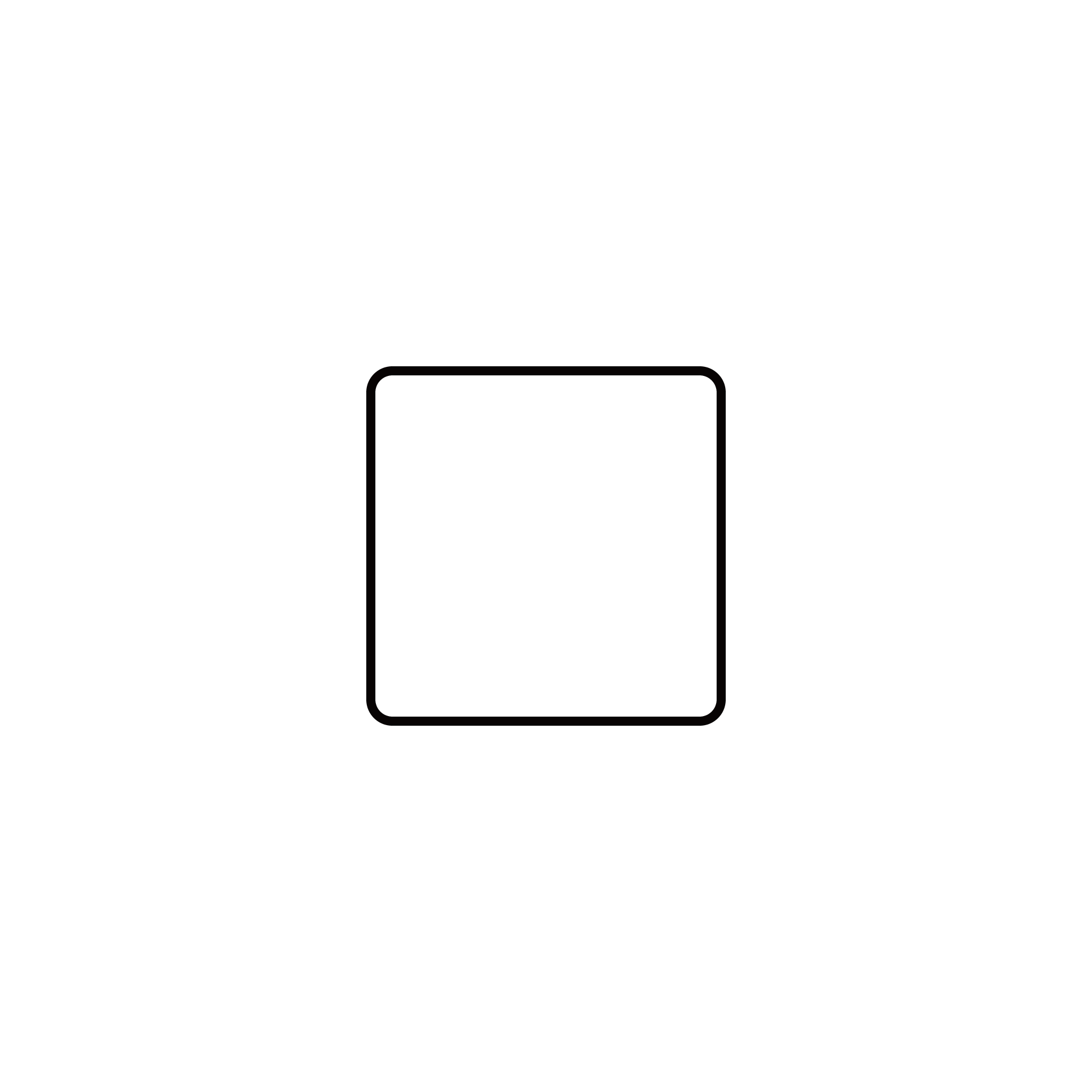}}
            \hspace{0.1cm}
            \subfigure[\label{fig:simulationEnv_PE2}]
            {\includegraphics[width=0.2\textwidth]{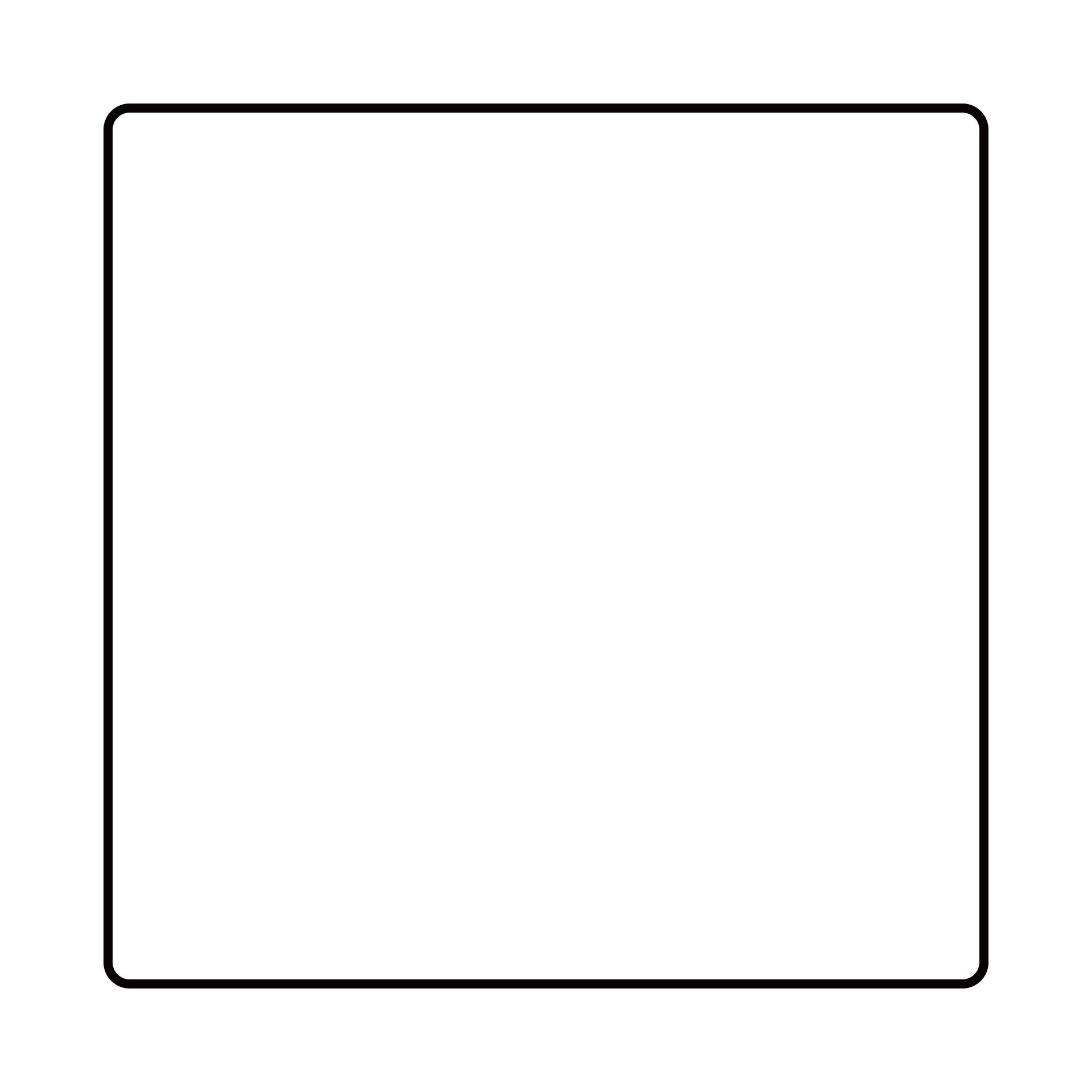}}
            \par
            \vspace{-2mm}
            \subfigure[\label{fig:simulationEnv_PE3}]
            {\includegraphics[width=0.2\textwidth]{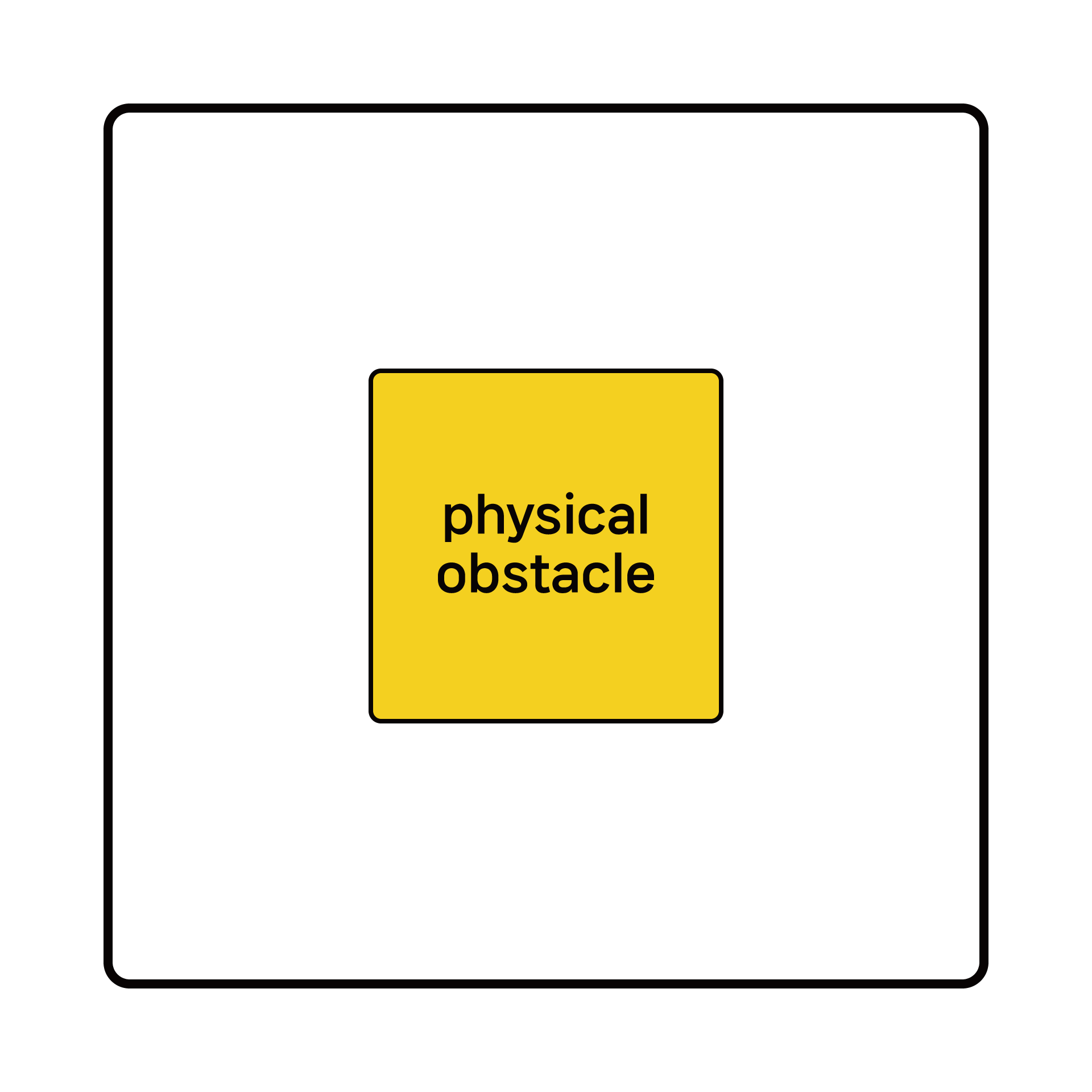}}
            \hspace{0.1cm}
            \subfigure[\label{fig:simulationEnv_PE4}]
            {\includegraphics[width=0.2\textwidth]{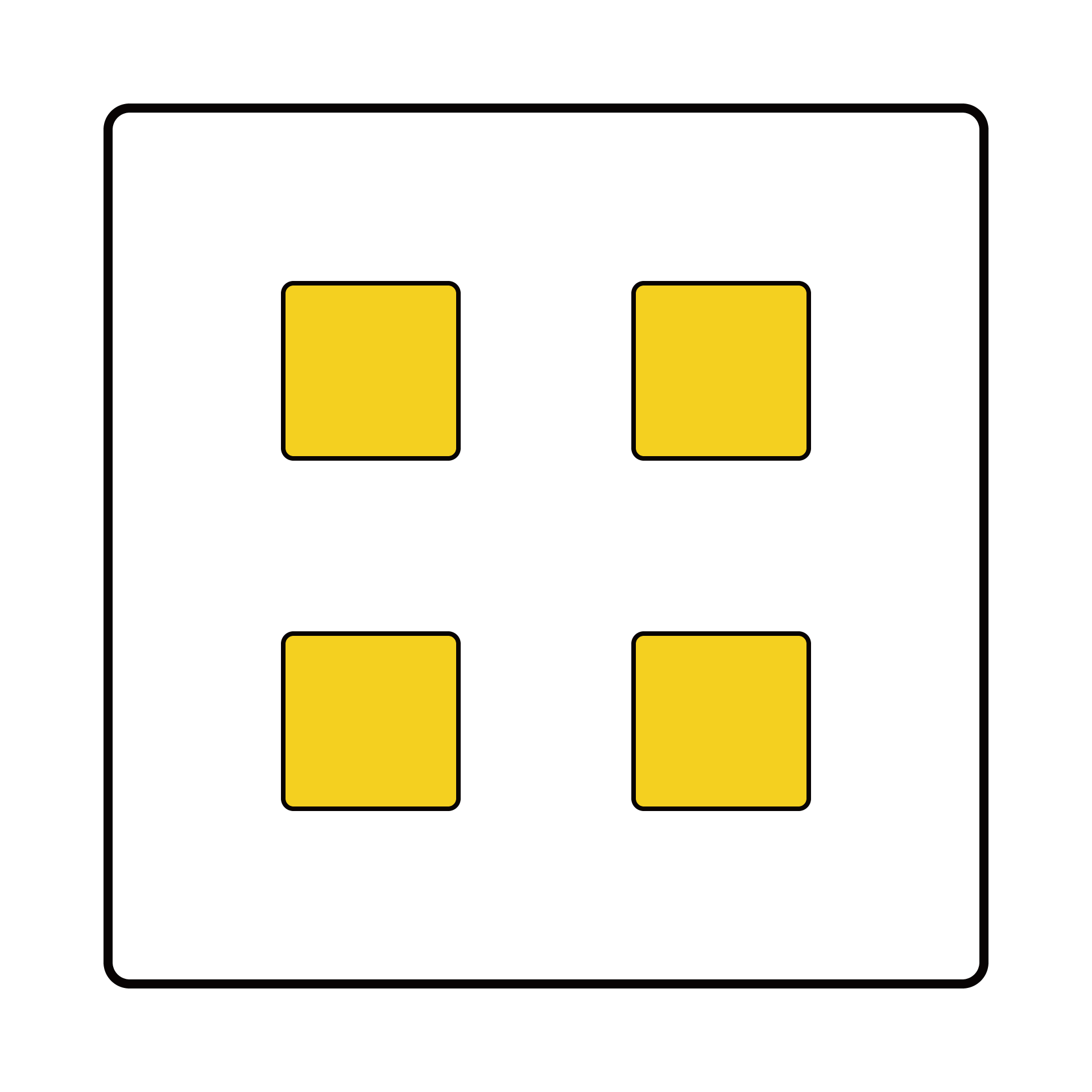}}
        \vspace{-2mm}
        \caption{Diagram of physical spaces used in our simulation tests: (a) an empty $4m \times 4m$ squared space used in \textbf{E1}. (b) an empty $10m \times 10m$ squared space used in \textbf{E2}. (c) a $10m \times 10m$ squared space with a single $4m \times 4m$-sized physical obstacle used in \textbf{E3}. (d) a $10m \times 10m$ squared space with four $2m \times 2m$-sized physical obstacles used in \textbf{E4}.}
        \label{fig:simulationEnv_PE}
        \vspace{-4mm}
    \end{figure}

\subsubsection{F-ARC}
The ARC algorithm introduced by Williams et al. [\citeyear{williams2021arc}] aims to provide passive haptic to users and redirects a user by considering spatial alignment, which is calculated as a function of the distances from a user to obstacles in the virtual and physical space, respectively (function $ComputeMisalign$ in Algorithm \ref{alg:F-ARC}). Specifically, ARC adjusts the translation gain according to the difference between two distances; distance from the user and an obstacle in the front in the virtual and physical space, respectively. Moreover, ARC adjusts the curvature gain in the direction with less difference between such distances among leftwards and rightwards. Lastly, ARC adjusts the rotation gain by comparing the current and previous frames' spatial alignment if a user rotates in place.

Alternatively, F-ARC determines the final translation gain and curvature gain by taking a weighted average of the gains computed from the spatial alignment of the current and future position, as shown in Figure \ref{fig:F-ARC} (c).
It also utilizes the future frame information instead of previous frame information when calculating the rotation gain (Algorithm \ref{alg:F-ARC}). In other words, F-RDW adjusts the level of the rotation gain after comparing the current and future frame's spatial alignment. Although the user's body direction in the physical and virtual space can be an essential attribute when calculating the spatial alignment at a stricter level, we did not include those values as the output scope of our prediction model in this study. Therefore, we consider the direction of the user's body in the virtual future position as the direction of the vector between the physical current position and the future position overlaid in the physical space.

    \begin{figure*}[t]
        \centering
        \includegraphics[width=0.9\textwidth]{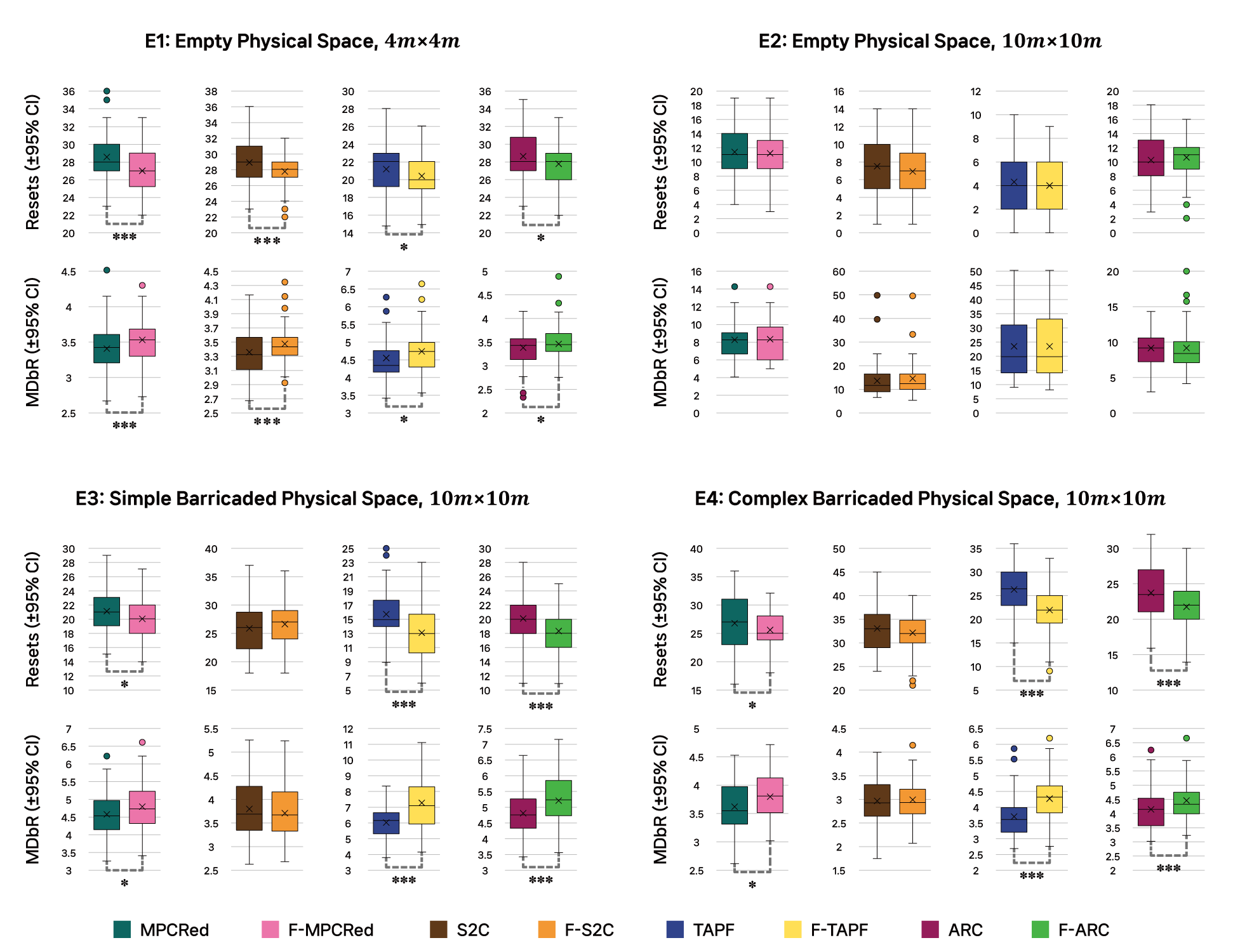}
        \vspace{-2mm}
        \caption{Visualization of the results of \textbf{E1-E4} in main simulation tests.}
        \label{fig:E1E4_Results}
    \vspace{-2mm}
    \end{figure*}

\section{Experimental Evaluation}
\subsection{Simulation Test Design}
We conducted simulation tests with Monte Carlo simulation to analyze the effects of integrating our proposed mechanism with various existing RDW methods. Thus, the independent variable of our experiments \textbf{E1-E4} is whether or not we apply our proposed mechanism to the vanilla RDW method, and the dependent variables are the RDW metrics: the number of reset occurrences and the mean distance between resets in the virtual environment (MDbR). 

We implemented the original RDW methods -- MPCRed, S2C, TAPF, and ARC -- and their corresponding F-RDW methods on OpenRDW \cite{li2021openrdw}. We also empirically determined the $\mu$ value (ratio of the current weight to the future weight) for F-S2C, F-TAPF, and F-ARC, to 0.5, 0.7, and 0.5, respectively; please refer to our ablation study in Section 5.1 for details on how we conducted the experiment for determining the $\mu$ value.
We then configured four physical spaces as depicted in Figure \ref{fig:simulationEnv_PE}, and we simulated each RDW method in the virtual environment as shown in Figure \ref{fig:simulationEnv_VE}. We designed the simulated users to collect targets generated within $0.2m \sim 8.0m$ of a radius around themselves. These targets were generated until simulated users traveled approximately 100$m$ in total. We also used the same constant parameter values and the same detection threshold used in initial studies of each RDW method \cite{nescher2014planning, razzaque2005redirected, thomas2019general, williams2021arc}.
Moreover, we initialized the user's position and orientation in the physical and virtual space with random values at the beginning of each trial \cite{hirt2022chaotic}.

For each vanilla RDW method and F-RDW method pair, we collected the simulated data for 100 iterations. Then, we conducted a normality test (Shapiro-Wilk test) and an equal variance test (Levene's test). Depending on their results, we also conducted the Wilcoxon signed-rank test and t-test, if necessary. Moreover, we used the default values in OpenRDW \cite{li2021openrdw} when configuring other settings for the user, such as the translation speed, rotation speed, and the radius of a user body, for determining the reset trigger.

\vspace{-2mm}
\subsection{Result of the Main Simulation Test}
We conducted each simulation test \textbf{E1-E4} in a physical and virtual space, as shown in Figure \ref{fig:simulationEnv_PE} and Figure \ref{fig:simulationEnv_VE}, respectively.
Except \textbf{E2}, we confirmed that F-RDW methods significantly reduce the number of resets and increase MDbR compared to its original RDW method as shown in Figure \ref{fig:E1E4_Results}.

\subsection{Discussion for the Main Simulation Test}
We believe that the superiority of F-RDW in \textbf{E1} mainly originates from the size of the physical space that it is employed; since resets inevitably occur frequently in small spaces, any small changes in the redirection policy that were based on future information will be likely to enhance the performance.

While we discovered a statistically significant positive effect of our mechanism in \textbf{E1}, we did not find the same effect in \textbf{E2}, which was still conducted in an empty room but had a relatively larger physical space (Table \ref{table:SimulationTestTable}). We speculate that this is because the importance of future information decrease as the physical space increases (even in empty space) since there is less likelihood of a user approaching the physical boundary. In other words, we expect that there will be more opportunities to take advantage of future information for redirection algorithms in smaller physical spaces.

It is also remarkable that we discovered a positive effect of our mechanism in \textbf{E3} where we newly added a large obstacle to the environment used in \textbf{E2}. We think this is due to the idea that given a physical space with a fixed area, it is more challenging to redirect in spaces with an internal obstacle; hence exploiting future information would be helpful to redirect and evade those obstacles.

We also identified a larger positive effect of our mechanism in \textbf{E4} than in \textbf{E3}
(Table \ref{table:SimulationTestTable}). We believe that given a physical space with a fixed area, there exist more occasions to employ future information effectively in the redirection algorithm if there are more obstacles and complex configurations in the given space.

Lastly, in the case of S2C, we identified that F-S2C outperforms the vanilla S2C in \textbf{E1}, but we did not find any statistically significant difference between those two algorithms in \textbf{E2-E4}. For \textbf{E2}, the results seem to arise from the same reason with other F-RDW methods as discussed above. For \textbf{E3-E4}, we believe that F-S2C may not have as many chances as other F-RDW methods in taking advantage of future information due to its internal mechanism that doesn't consider the existence of physical obstacles; note that F-RDW respects the internal mechanism of its base RDW method.
            
\subsection{User Study Design}
Extending our simulation test, we conducted a user study to examine the effectiveness of our mechanism in real-world environments as well. In our user study, we compared TAPF and F-TAPF, which showed the best performances in each group of our simulation tests.

We also recruited 28 participants (male: female = 4.6: 1, the average age = 25.69, $SD$ = 4.52) with normal vision or normal corrected vision since the G* power program \cite{faul2007g} reported that a minimum of 28 participants are needed for our user study with significance level (= 0.05), medium effect size (= 0.25) (according to Cohen's guidelines), and statistical power (= 0.80). The average number of their previous VR experiences was 2.1 ($SD$ = 1.1).

    \begin{figure}[t]
        \centering
            \includegraphics[width=0.45\textwidth]{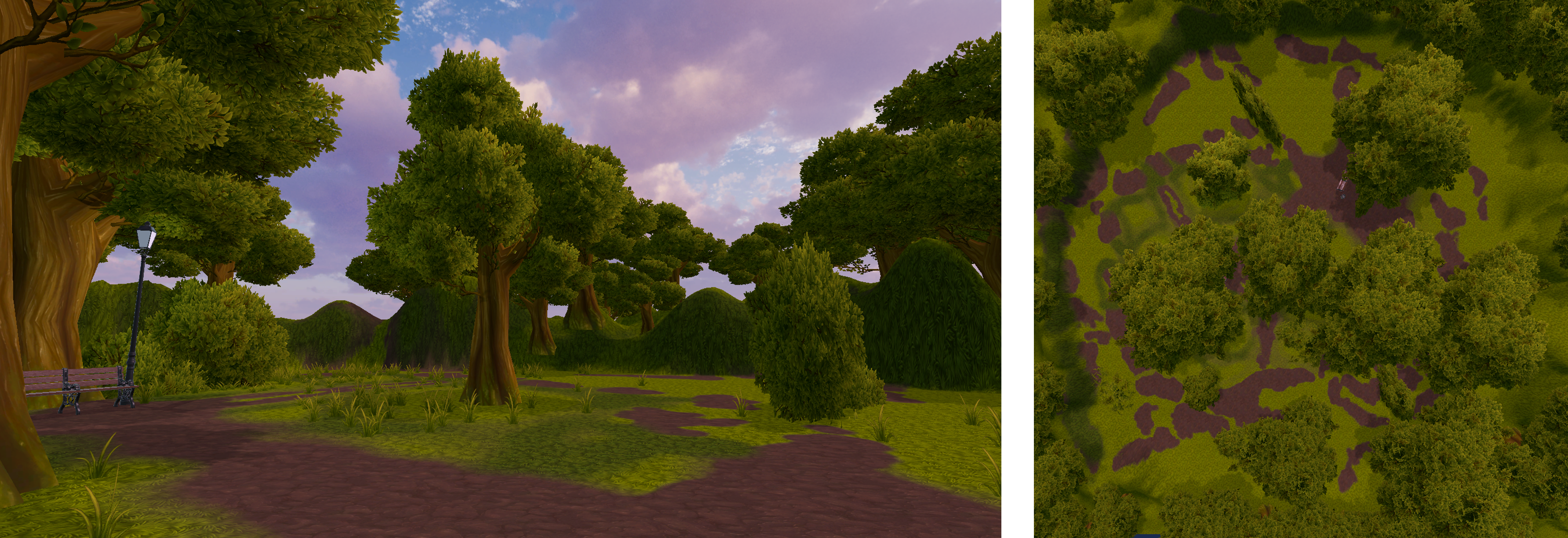}
        \vspace{-2mm}
        \caption{Virtual environment used in our user study. Participants were asked to collect garbage bags in a $50m \times 50m$ square-shaped park in the virtual environment.}
        \label{fig:UserStudyEnv_VE}
        \vspace{-5mm}
    \end{figure}

For our user study environment, we installed a wireless HTC VIVE Pro eye HMD and used four base stations for inside-out tracking in a physical space ($4m \times 4m$). The average FPS for our VR program in the hardware settings was 60 FPS.

During the experiment, participants engaged in a task to collect as many virtual targets (garbage bags) in a virtual park as shown in Figure 
\ref{fig:UserStudyEnv_VE}. 
Since users may explore in diverse path patterns, such as in mixed exploration \cite{jeon2022dynamic}, once a user collected a garbage bag, we generated the next garbage bag at a random location within a radius of $0.2m \sim 8.0m$ around the user that did not overlap with other virtual obstacles.

Before participants started their VR experience, we first explained the overview of our experiment and instructed the participants on how to put on an HMD and VIVE controller for 5 minutes. We also informed the participants that 10 US dollars (\$) would be given as a reward after completion. The next 5 minutes were given to the participants to experience the virtual environment (practice zone) with real walking. Afterward, participants filled out a pre-SSQ \cite{kennedy1993simulator} questionnaire during their 5 minutes of break time. 
Then, participants put on their VR devices and engaged in the VR task (3 minutes) twice, with 5 minutes for break time and filling out questionnaires after each iteration. Lastly, we interviewed each participant for 5 minutes. In order to prevent any possible carryover effect during our experiment, we also considered counter balancing. We used four types of questionnaires -- SSQ \cite{kennedy1993simulator}, and E2I \cite{lin2002effects} for measuring motion sickness, and immersion, respectively, and SUSPQ \cite{usoh2000using}, and IPQ \cite{schubert2001experience} for measuring presence -- with a 7-point Likert scale when collecting user data.

\vspace{-1mm}
\subsection{Result of E5: User Study}
\begin{figure}[t]
        \centering
        \includegraphics[width=0.47\textwidth]{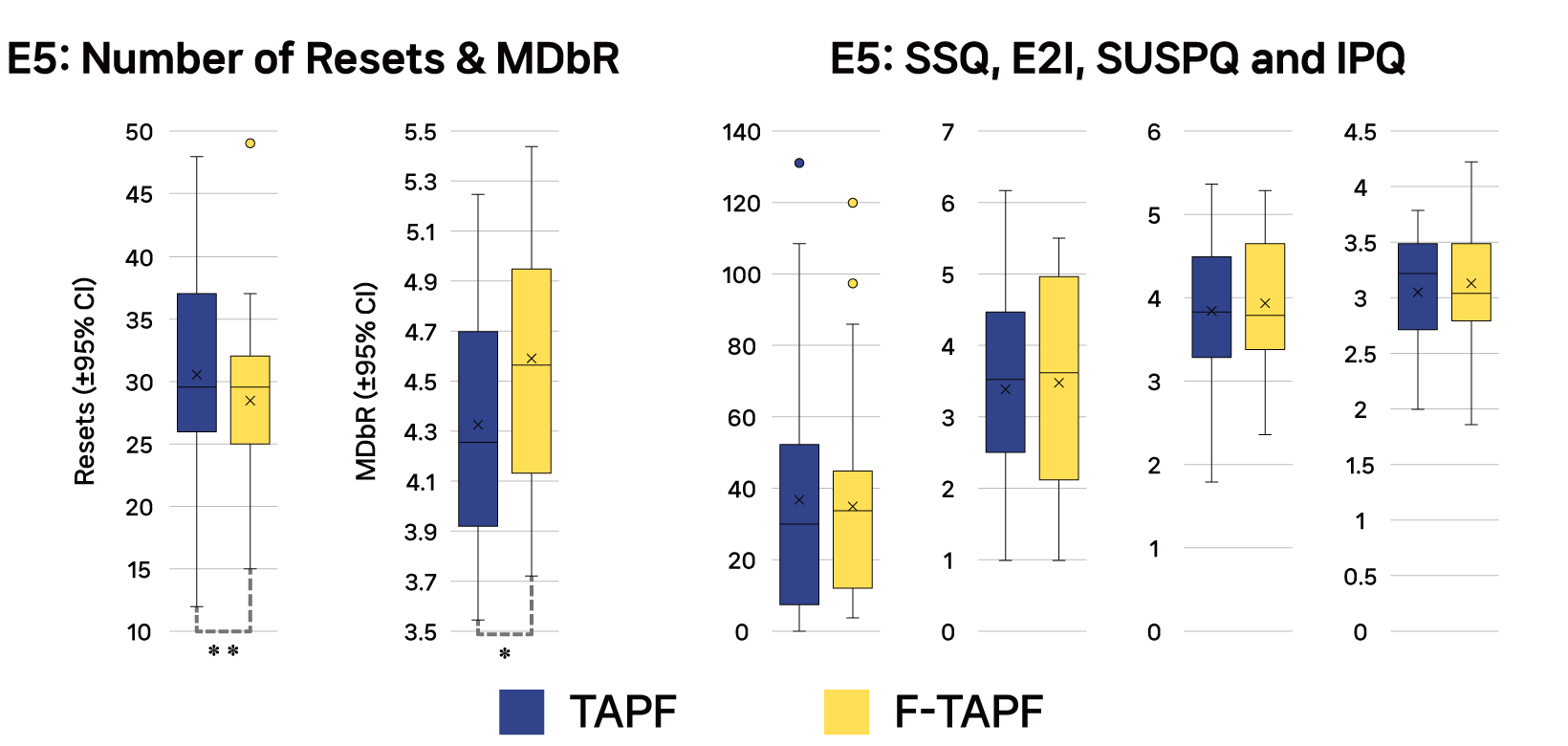}
        \caption{Visualization of the results of \textbf{E5} in our user study.}
        \label{fig:E5_Result}
        \vspace{-5mm}
    \end{figure}

As illustrated in Figure \ref{fig:E5_Result}, F-TAPF outperformed TAPF in both metrics: the number of resets $(z(26) = 2.535, p = 0.005)$ and MDbR $(t(26) = 2.27, p = 0.02)$. Our experiment validated that F-TAPF shows better redirection performance than the original TAPF algorithm.

In the meantime, there was no statistically significant difference between the mean values of the Likert scores obtained from the four questionnaires pertaining to TAPF and F-TAPF, although the scores for F-TAPF were slightly better than TAPF. We expect to identify stronger cues from the questionnaire results if the future prediction accuracy of F-RDW improves or a sophisticated application strategy for F-RDW is integrated.

    \begin{figure*}[t]
        \centering
        \includegraphics[width=1.0\textwidth]{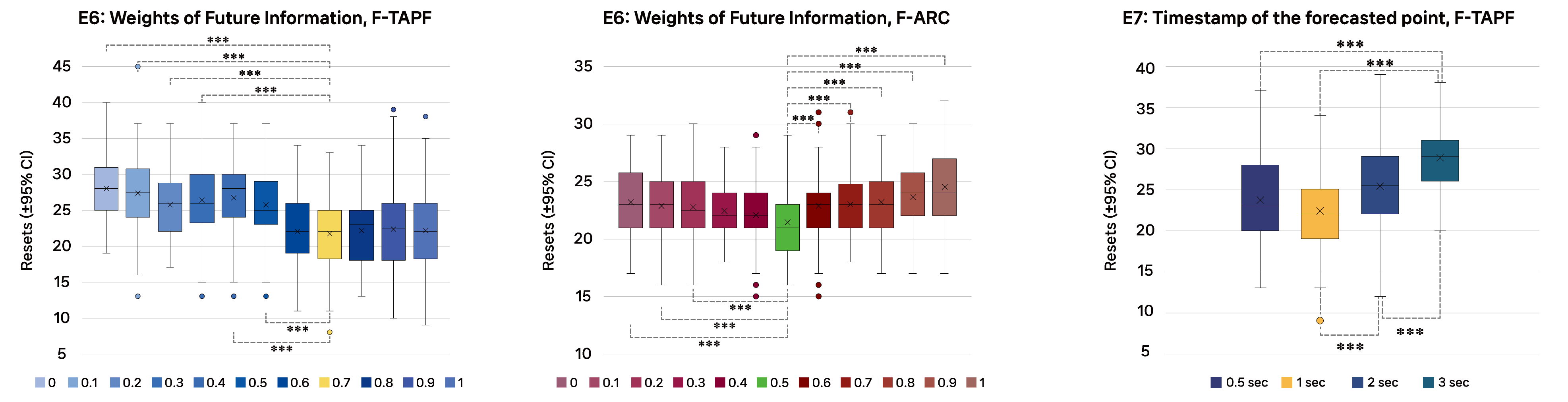}
        \caption{Bar plots of the number of resets in \textbf{E6-E7} conducted for ablation studies. 
        Asterisks indicate the level of significance with respect to each optimal value in \textbf{E6}.}
        \label{fig:AblationStudy_futureRatio_futureTimestamp}
    \vspace{-2mm}
    \end{figure*}

\section{Ablation Study}
\subsection{Weights of Current and Future Information}
Prior to our main simulation tests, we measured the change in the redirection performance when the weights of the current and future information $\mu$ were manipulated in F-RDW. Under the same condition with \textbf{E4}, we conducted \textbf{E6} on F-S2C, F-TAPF, and F-ARC to analyze such a relation.
The results of Kruskal-Wallis test demonstrated that $\mu_{F-TAPF}$ $({\chi}_{10,1089}^2=215.751,p < .001,\eta^2_p = .19)$
 has a large effect on the number of resets and $\mu_{F-ARC}$ $({\chi}_{10,1089}^2=73.745,p < .001,\eta^2_p = .059)$
 has a small effect on the number of resets. Based on the results of Mann-Whitney U-test for post-hoc analysis with a p-value from the Bonferroni correction ($p_c = .0009$) (Figure \ref{fig:AblationStudy_futureRatio_futureTimestamp}), we selected $\mu_{F-TAPF}=0.7$ and $\mu_{F-ARC}=0.5$, which showed the lowest average number of resets in the main simulation test. F-TAPF seemed to show better redirection performance with bigger future weights, while F-ARC showed the lowest average number of resets when the current and future weights were equal.
 However, F-S2C did not show a peak performance at a particular $\mu$ value, so we set $\mu_{F-S2C}=0.5$ and conducted our main simulation tests for further analysis.
 Moreover, although we used a static $\mu$ value for each RDW method in this study, we believe that using a dynamic $\mu$ value depending on the context can be more powerful. For instance, if a user approaches closely to the physical boundary, the current information may be (possibly) more important than the future information.

\subsection{Timestamp of the Forecasted Point}
We also examined how the redirection performance changes depending on the timestamp of the forecasted point ($F_t$) that our LSTM-based prediction model outputs, when training the prediction model before integrating it with the main F-RDW mechanism and initiating our main simulation test.
To examine such a relation, we experimented \textbf{E7} on F-TAPF under the same environment with \textbf{E4}. The results of Kruskal-Wallis test demonstrated that $F_t$ (${\chi}_{3,396}^2=84.159,p < .001,\eta^2_p = .2)$ has a large effect on the number of resets. Based on the results of Mann-Whitney U-test for post-hoc analysis with a p-value from the Bonferroni correction ($p_c = .0008$) (Figure \ref{fig:AblationStudy_futureRatio_futureTimestamp}), we selected $F_t=1$ which showed the lowest average number of resets in the main simulation test. However, we acknowledge that different optimal $F_t$ values may exist for different user conditions. For example, a dynamic $F_t$ that changes according to the user's velocity can be more desirable, rather than using static $F_t$ values.

    \begin{table}[t]
        \centering
        \caption{Performance of F-TAPF (\textbf{E7}) with different neural networks when employed in the same environment used in \textbf{E4}.
        For each neural network, we used the same hyperparameter values (Table \ref{table:hyperparameters1}) that we used for training the LSTM network of our prediction model. 
        }
        \vspace{-2mm}
        {
        \begin{tabular}{lccccc}
            \cline{1-5}
            Architecture & MDE ($m$) & SD ($m$) & Reset (\#) & MDbR ($m$) &  \\\cline{1-5}
            LSTM & \textbf{0.45} & 0.35 & \textbf{21.97} & \textbf{4.27} &  \\ 
            GRU & 0.48 & 0.33 & 22.32 & 4.18 &  \\
            Transformer & 0.47 & 0.32 & 22.16 & 4.22 &  \\\cline{1-5}
        \end{tabular}
        }
        \label{table:AblationStudy_Architecture}
        \vspace{-3mm}
    \end{table}

\vspace{0mm}
\subsection{Architectures for Forecasting Future Position}
Bremer et al. [\citeyear{bremer2021predicting}] demonstrate that LSTM \cite{hochreiter1997long} is a better architecture for the neural network than GRU \cite{cho2014learning} when predicting the future position of a user with spatial data and eye-tracking data. Moreover, our analysis in \textbf{E7} demonstrated that LSTM-based model shows better performance than Transformer-based model \cite{vaswani2017attention}, which is generally known to perform very well in time-series forecasting (Table \ref{table:AblationStudy_Architecture}). This is because Transformer networks generally require large-scale training dataset \cite{dosovitskiy2020image}. Therefore, we expect that acquiring more dataset to fuel the Transformer architecture could further enhance the performance of F-RDW methods.

\vspace{0mm}        
\section{Limitation}
First, although we applied our mechanism to the \emph{subtle} technique of various RDW methods, it is essentially difficult to directly reflect our mechanism in the \emph{reset} technique of each RDW method. This is because it is difficult to extract features from eye-tracking data when a reset technique, which utilizes rotation gain, is initiated. Thus, this leads to poor prediction accuracy during resets.
Second, as discovered in \textbf{E2}, there are fewer opportunities for F-RDW to exercise its capacity in physical environments with large walkable spaces since it is obvious that the number of reset occurrences decreases in such conditions. However, we did verify that F-RDW does not have any negative effect even in those environments.
Lastly, we limited the time for each episode in our user study to 3 minutes in order to prevent any carryover effect that might arise in within-subject design experiments. Therefore, we expect that conducting our experiment in much more extensive environments with more various metrics \cite{azmandian2022validating,williams2022eni} will further augment the robustness of F-RDW.

\vspace{-1mm}
\section{Conclusion and Future Work}
Predictive RDW methods that exploit users' future information reduce the number of reset occurrences that degrade users' sense of immersion and presence. However, poor prediction accuracy of the future information leads to poor redirection performance, and achieving real-time serving is infeasible due to the massive computation cost for considering a nearly infinite number of possible future spatial information. To bypass these obstacles, previous studies depended on additional constraints on the virtual environment's layout or the user's walking direction. 
To tackle such challenges at a more fundamental level, we propose a novel predictive redirection mechanism (F-RDW). By using machine learning, F-RDW is independent of any constraints or assumptions when predicting the future position of a user in the virtual space and reflects the predicted values into existing RDW methods with appropriate strategies. From our simulation tests and user study, we validated that F-RDW improves the redirection performance of existing RDW methods in small physical spaces or complex environments with obstacles.

For our future work, we suggest using more various types of multi-modal data (e.g., audio and body pose data) as it is known to augment the prediction accuracy when predicting the user's future information \cite{islam2021cybersickness,pakdamanian2021deeptake}. Multi-modal fusion neural networks may potentially yield further improvements by understanding the user's purpose behind their travel behavior or by compensating for the poor prediction accuracy during resets.
Second, we plan to study how to acquire the optimal hyperparameter values of F-RDW -- $\mu$ and $F_t$ -- dynamically. Further investigation on how to efficiently reflect the characteristics of users (e.g., path pattern), base RDW method, and the employed environment in determining those values would allow additional performance enhancement of F-RDW.

\begin{acks}
This research was supported by the National Research Foundation of Korea(NRF) grant funded by the Korea government(MSIT). (No. NRF-2020R1A2C2014622)
\end{acks}

\bibliographystyle{ACM-Reference-Format}
\bibliography{reference}


\end{document}